\documentclass{article}

\PassOptionsToPackage{numbers, compress}{natbib}
 \usepackage[preprint]{neurips_2026}

\usepackage{natbib}
\usepackage[utf8]{inputenc} 
\usepackage[T1]{fontenc}    
\usepackage{hyperref}       
\usepackage{url}            
\usepackage{booktabs}       
\usepackage{amsfonts}       
\usepackage{nicefrac}       
\usepackage{microtype}      
\usepackage{xcolor}         
\usepackage{microtype}
\usepackage{graphicx}
\usepackage{subcaption}
\usepackage{multirow}
\usepackage{booktabs} 
\usepackage{tcolorbox}
\tcbuselibrary{breakable}
\usepackage{enumitem}
\usepackage{hyperref}
\usepackage{amsmath}
\usepackage{alltt}
\usepackage{tabularx}
\usepackage{wrapfig}

\newtheorem{theorem}{Theorem}[section]

\usepackage[table]{xcolor}
\definecolor{lightyellow}{RGB}{255, 249, 196}
\usepackage[textsize=tiny]{todonotes}
\newcommand{\name}{ImageAuditor}
\newcommand{\rag}{IRAG}
\newcommand{\tprfpr}{TPR$_{5\%}$}
\newcommand{\algo}{RGPO}

\title{{\name}: Membership Inference Attack against Image-based Retrieval-Augmented Generation}

\author{
{\bfseries Jinghuai Zhang\textsuperscript{1} \quad Pengyue Yu\textsuperscript{1} \quad Zhexiao Lin\textsuperscript{2} \quad Kunlin Cai\textsuperscript{1}}\\[3pt]
{\bfseries Fnu Suya\textsuperscript{3} \bfseries Yuan Tian\textsuperscript{1}}\\[3pt]
\textsuperscript{1}UCLA \quad \textsuperscript{2}University of California, Berkeley \quad \textsuperscript{3}University of Tennessee, Knoxville \\[4pt]
\texttt{\{jinghuai1998,pengyue2024,kunlin96,yuant\}@g.ucla.edu}\\[2pt]
\texttt{zhexiaolin@berkeley.edu} \quad \texttt{fsuya@utk.edu} \\
}

\begin{document}

\maketitle

 \begin{abstract}
Image-based Retrieval-Augmented Generation (IRAG) conditions a frozen generator on reference images retrieved from an external database, supporting both text-to-image (T2I) and question answering (Q\&A) tasks. Because these databases are opaque and typically web-scraped, copyright holders need ways to audit whether specific images appear in them. 
While prior work employs membership inference attacks (MIAs) to audit uni-modal, text-based RAG, they fail to transfer to IRAG due to \textit{two} fundamental challenges. First, \textit{cross-modal retrieval}: text-RAG MIAs force retrieval of the target passage by injecting its content into the query, which is unavailable in IRAG since images cannot be embedded into text queries; even accurate image captions fail to bridge the modality gap. Second, \textit{discriminative signal extraction}: text-RAG MIAs extract membership signals by prompting the generator to answer multiple questions over the target passage, whereas T2I generators in IRAG produce images rather than follow Q\&A commands.
To fill this gap, we introduce the first MIA tailored to IRAG, \emph{ImageAuditor}, which decomposes each attack query into a retrieval segment and an extraction segment, enabling dedicated optimization for each challenge.
For retrieval, we propose Reward-Guided Policy Optimization (RGPO), which updates a stochastic policy from reward-ranked candidates to navigate the cross-modal embedding landscape and admits finite-sample optimality guarantees to balance exploration and exploitation.
For extraction, we analyze the distribution of the MIA score to guide the co-design of the prompting strategy and scoring rule, and derive task-specific instantiations for T2I and Q\&A tasks. 
We aggregate signals across queries via K-means clustering for reliable membership decisions. Across IRAG systems built on SDXL, SD1.5, Kandinsky, LLaVA-1.6, and Qwen2.5-VL, ImageAuditor exceeds 80\% AUROC with only four queries per audited image and remains robust across diverse settings.

\end{abstract}
\section{Introduction}

\begin{wrapfigure}{r}{0.4\textwidth}
    \centering
    \includegraphics[width=0.4\textwidth]{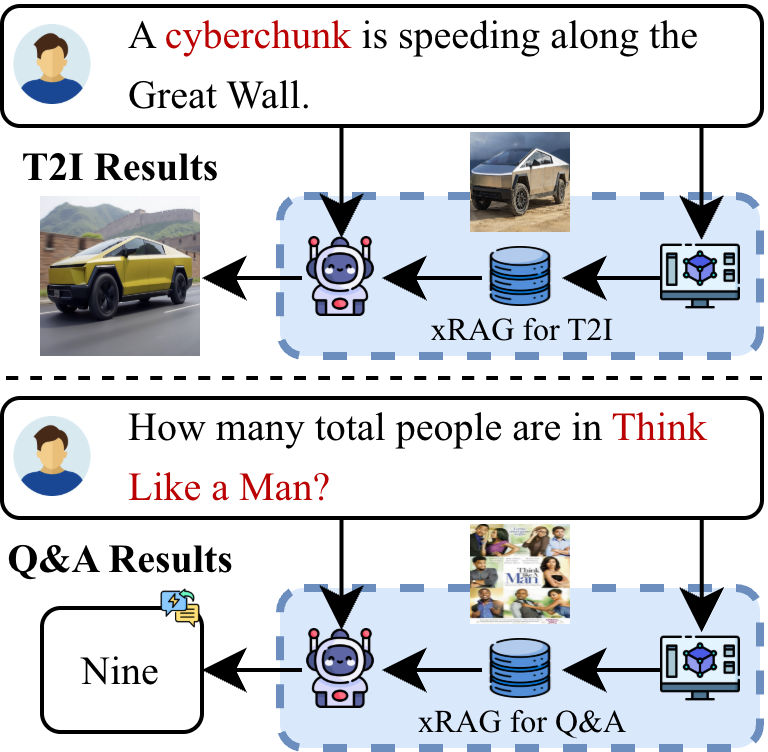}
    \caption{{\rag} employs an external image database to generate faithful images (top) and produce reliable answers to knowledge-dependent queries (bottom).}
    \label{fig:teasar}
\end{wrapfigure}

Image-based Retrieval-Augmented Generation ({\rag})~\citep{lyu2025realrag,chen2024mllm} grounds generative models in external visual knowledge by retrieving reference images from an image database. It has gained increasing popularity in two major multimodal generation tasks (Figure~\ref{fig:teasar}): text-to-image generation (T2I)~\citep{lyu2025realrag,shalev2025imagerag,chen2022re}, where retrieved images supply concepts, identities, and styles beyond the reach of base diffusion models~\citep{podell2023sdxl}; and question answering (Q\&A)~\citep{chen2024mllm,chen2022murag,lin2022retrieval}, where retrieved images provide knowledge-dependent visual evidence for vision-language models~\citep{wang2024qwen2}. For consistency, this paper focuses on {\rag} systems that take text as input and produce either images (T2I) or text (Q\&A).

This rapid adoption also exposes a privacy and copyright surface~\citep{luo2025imagesentinel,chen2025safeguarding}. Because these databases are typically web-crawled with poorly documented provenance, copyright holders (e.g., artists), dataset owners, and privacy regulators have no reliable way to verify whether their images have been silently incorporated into a deployed system, motivating the need for principled auditing tools.

Membership inference attacks (MIA)~\citep{shokri2017membership}, which infer whether a record belongs to a target system, provide a natural auditing primitive. Traditional MIAs~\citep{shokri2017membership,carlini2022membership} target a model's training data by exploiting memorization signals in its weights, and therefore do not apply when membership resides in an external database. Recent text-RAG MIAs~\citep{naseh2025riddle,li2025generating,liu2025mask} probe the database via question--answer interactions, but they implicitly assume \emph{single-modal retrieval}, where the attacker can inject the target document into the query to force retrieval, and \emph{text-format outputs}, where responses can be matched against pre-computed answers. Both assumptions break down in {\rag} settings. A text query cannot carry an image target, and we show in Section~\ref{sec:exp_main} that even substituting a strong caption fails to bridge the modality gap. Furthermore, {\rag} for T2I generation emits images rather than text and does not explicitly answer the query, leaving Q\&A-style probing without a signal to match. We formalize these obstacles as \emph{two fundamental challenges} in Section~\ref{sec:challenge}.



\noindent
\textbf{Contributions} To fill this gap, we introduce \textit{\name}, the first MIA tailored to {\rag}, operating in a realistic setting with only a handful of attack queries per audited image (Section~\ref{sec:threat-model}). Our key design decomposes each query into two functionally isolated segments: a \emph{retrieval segment} that brings the test image into the generator's context when present, and an \emph{extraction segment} that elicits a discriminative MIA signal from the resulting output. This decomposition allows each segment to be explicitly optimized against a clean and dedicated objective, avoiding conflicts inherent in coupled optimization.



For the retrieval segment, a natural starting point is the suffix-optimization toolkit developed for RAG \emph{poisoning} attacks~\citep{zou2025poisonedrag,xue2024badrag}, which consider a structurally similar problem of crafting text that retrieves a chosen target. However, their gradient-based optimizers are designed for single-modal text retrieval and are easily trapped in local minima on the non-smooth cross-modal embedding landscape, motivating a gradient-free, exploration-heavy alternative. We therefore propose \textit{Reward-Guided Policy Optimization} ({\algo}), which updates a stochastic token policy from reward-ranked candidates to balance exploration and exploitation, and comes with a finite-sample optimality guarantee (Theorem~\ref{thm:rgpo} in Section~\ref{sec:ret}). 

For the extraction segment, we derive a mixture-identity decomposition of the MIA score that cleanly separates retrieval probability from the extraction hit/miss gap, and use it to guide task-specific instantiations for T2I and Q\&A (Section~\ref{sec:ext}). We further pair these instantiations with K-means aggregation across queries (Section~\ref{sec:pipeline}). Across {\rag} systems built on six generators and diverse datasets, {\name} consistently outperforms adapted baselines, exceeding 80\% AUROC and reaching around 50\% TPR@5\%FPR with only four queries per audited image. These values are well above the bar for a strong MIA~\citep{carlini2022membership}, establishing {\name} as a practical auditing tool.

\section{Background and Threat Model}
\subsection{Background}
\noindent
\textbf{{\rag}.}
We describe an {\rag} system as a tuple $S = (R, D, G)$ consisting of a retriever $R$, an image database $D$, and a generator $G$, where each image $I \in D$ is encoded by an embedding model $E$ that $R$ uses for retrieval. The key idea is to leverage external image knowledge (e.g., web-crawled images) to compensate for knowledge gaps in the generator's parameters.  In this work, we cover {\rag} for both T2I and Q\&A tasks. Formally, given a text query $T$, the system retrieves the top-$K$ relevant images and produces output $y$ (an image for T2I, a text answer for Q\&A).

\noindent
\textbf{Membership inference attack (MIA).} MIA~\cite{shokri2017membership}, which aims to determine whether a record belongs to a target database, is a prominent framework for auditing data privacy in ML systems. Prior MIAs~\cite{carlini2022membership,he2025towards,pang2023black} target a model's training data and do not transfer to {\rag} systems, where the target data must first be retrieved before it can influence the output. While recent work~\cite{liu2025mask,naseh2025riddle} explores MIAs against text-based RAG, their query designs fail to reliably retrieve the target image due to the modality gap, especially under the small retrieval budgets used in {\rag} (see Sec.~\ref{sec:challenge}). Moreover, their Q\&A-style attack strategies do not transfer to {\rag} systems whose generator is a text-to-image model. A separate line of work~\cite{chen2025safeguarding,luo2025imagesentinel} addresses {\rag} copyright concerns from the \emph{defender's} side, embedding watermark patterns into images so that owners can later assert ownership. These approaches require proactive modification of the protected images and do not help general auditors (e.g., ordinary users) whose published data lacks such sentinels.

\subsection{Threat Model}\label{sec:threat-model}
\noindent
\textbf{Attacker's goal.} The adversary acts as a data auditor (e.g., a copyright holder, a database owner, or a privacy regulator). Given a test image $I$ and a target {\rag} system $S_{\text{tgt}}$, the attacker aims to infer the membership $m \in \{0,1\}$ of $I$ in the image database $D_{\text{tgt}}$. We frame this as a binary classification task (member vs.\ non-member) intended to detect unauthorized use of copyrighted images.

\noindent
\textbf{Attacker's capabilities.} We consider a realistic and challenging attack setting. The attacker can only interact with the target {\rag} system $S_{\text{tgt}}$ through its API, operate under a limited query budget (e.g., 4--5 per test image) to remain stealthy, and has no knowledge of the images in $D_{\text{tgt}}$. For transparency, {\rag} systems typically disclose the names of their AI components, such as the embedding model and the generator. Following the standard threat model in prior RAG attacks~\cite{xue2024badrag,zou2025poisonedrag,geng2025unic,wang2026joint,chaudhari2024phantom}, we assume by default that the attacker has access to $E_{\text{tgt}}$. This assumption is realistic because (i) deployed {\rag} systems overwhelmingly adopt off-the-shelf CLIP variants~\cite{radford2021learning} as their retrievers~\cite{luo2025imagesentinel,shalev2025imagerag,chen2025safeguarding,chen2024mllm,lyu2025realrag,chen2022re}, since training a custom cross-modal encoder is expensive and rarely improves retrieval quality; and (ii) commercial providers further expose embedding APIs that return query embeddings directly.\footnote{For example, Google's multimodal embedding model \texttt{gemini-embedding-2} is accessible via API.} Concretely, the attacker submits an image or text query and obtains its embedding via local deployment or the API. To stress-test this assumption, we additionally evaluate a restrictive scenario in which the attacker has no access to $E_{\text{tgt}}$ and instead relies on a weaker shadow embedding model $E_{\text{sh}}$ (e.g., a smaller model trained with a different architecture or on a different data distribution); {\name} remains effective in this setting (Sec.~\ref{sec:exp_abl}). Finally, the attacker has access to a captioning model and, optionally, a small shadow dataset drawn from the same domain as the test images, which is readily collected from the Internet.


\section{Method}
\begin{figure*}[t]
    \centering
    \includegraphics[width=0.96\textwidth]{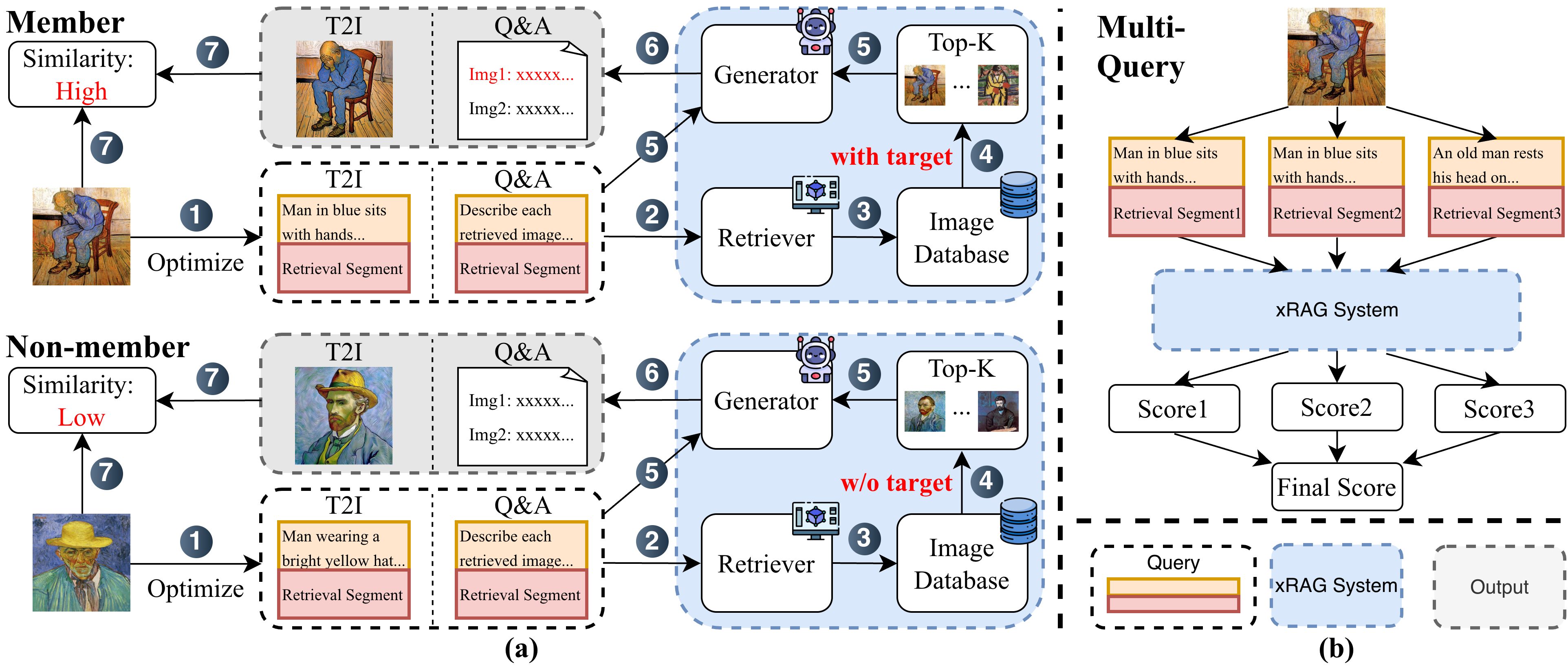}
    \caption{Framework overview. \textbf{(a) Single-query attack.} Given a test image, we construct a text query composed of an extraction segment (\textcolor{orange}{orange}) and a retrieval segment (\textcolor{red}{red}). The retrieval segment is optimized via RGPO (Sec.~\ref{sec:ret}) such that, when the test image lies in the database, the query retrieves it with high probability. The extraction segment and the similarity-based scoring rule are \textit{co-designed} (Sec.~\ref{sec:ext}) to elicit outputs that produce discriminative MIA signals: members yield outputs that closely resemble the test image (top), while non-members do not (bottom). \textbf{(b) Multi-query attack.} {\name} pairs multiple extraction and retrieval segments, scores each query, and aggregates the scores via K-means clustering for improved attack results. (Sec.~\ref{sec:pipeline}).}
    \label{fig:framework}
\end{figure*}

\subsection{Key Challenges}
\label{sec:challenge}

Designing an effective MIA against {\rag} faces two fundamental challenges. \textbf{Challenge 1 (C1): Cross-modal retrieval.} Unlike traditional MIAs~\citep{shokri2017membership,carlini2022membership}, where members are memorized in model parameters and can be directly probed, our setting requires the test image to be retrieved into the generation context before it can influence the output of $S_{\text{tgt}}$. Crafting a text query that reliably retrieves a target image is non-trivial: text tokens are discrete, the embedding space is cross-modal, and the retrieval budget is small (e.g., 1--3). Prior text-RAG MIAs~\citep{naseh2025riddle,li2025generating,liu2025mask} sidestep this by injecting the target passage directly into the query---an option unavailable here, since an image cannot be embedded into a text query, and even a strong caption fails to bridge the modality gap (Sec.~4.2). White-box suffix-optimization attacks~\citep{xue2024badrag,zou2025poisonedrag} also suffer, as they are easily trapped in local minima on the non-smooth cross-modal landscape. \textbf{Challenge 2 (C2): Discriminative signal extraction.} Even when retrieval succeeds, the attacker must still coerce the black-box generator into producing outputs whose distributions diverge between members and non-members. This is non-trivial across modalities: T2I generators produce unstructured images rather than formatted answers, while Q\&A generators tend to summarize multiple references and dilute the per-image signal. Prior Q\&A-style strategies~\citep{naseh2025riddle,liu2025mask} match generated text against pre-computed answers and require many queries per sample, which is incompatible with T2I outputs and impractical under realistic stealth budgets.

\textbf{Our solution: query decomposition.} {\name} addresses both challenges by decomposing each attack query into two functionally isolated segments---an extraction segment $T_{\text{ext}}$ and a retrieval segment $T_{\text{ret}}$---such that $T = T_{\text{ext}} \oplus T_{\text{ret}}$. The retrieval segment is optimized so that $I$ is retrieved with high probability whenever it lies in the database (addressing C1), while the extraction segment and the corresponding scoring rule are \emph{co-designed} from a principled analysis to elicit outputs that yield a discriminative member/non-member signal (addressing C2). This decomposition lets each component target a clean, dedicated objective. Below, we describe the retrieval segment, the extraction segment, and the score-aggregation strategy for multi-query attacks (see framework overview in Figure~\ref{fig:framework}).

\subsection{Retrieval Segment Optimization}
\label{sec:ret}

The retrieval segment aims to ensure that, given a designed extraction segment $T_{\text{ext}}$ (illustrated in Sec.~\ref{sec:ext}), the target system $S_{\text{tgt}}$ retrieves the test image $I$ as a reference for the query $T = T_{\text{ext}} \oplus T_{\text{ret}}$ with high probability whenever $I$ lies in the database. To this end, we optimize a contrastive reward that is observable to the attacker:
\begin{align*}
f(T_{\text{ret}}; I, T_{\text{ext}}) := \mathrm{sim} \big(E(T_{\text{ext}} \oplus T_{\text{ret}}), E(I)\big)
- \frac{\alpha}{|\mathcal{S}|} \sum_{J \in \mathcal{S}} \mathrm{sim} \big(E(T_{\text{ext}} \oplus T_{\text{ret}}), E(J)\big),
\end{align*}
where $\mathcal{S}$ is an optional set of shadow images and $\alpha$ balances attraction to the test image against repulsion from the shadow images. When no shadow set is available, we simply set $\alpha = 0$. Including a shadow set shapes a more favorable and generalizable optimization landscape, particularly under restrictive settings (see Table~\ref{tab:exp_shadow}). Maximizing the contrastive reward encourages the composed query to be close to the test image while remaining separated from generic distractors, which better approximates the retrieval event than plain similarity alone. Here $E$ denotes the embedding model accessible to the attacker (i.e., $E = E_{\text{tgt}}$ or $E = E_{\text{sh}}$, as introduced in the threat model).

A natural approach is to adapt optimization techniques from poisoning attacks on text-RAG (e.g., BadRAG~\cite{xue2024badrag}, PoisonedRAG~\cite{zou2025poisonedrag}). However, these methods suffer from two key limitations. (1) Gradient-based methods are sensitive to initialization and prone to poor local optima in the non-convex landscape induced by discrete token search, and the text-image modality gap further exacerbates this issue. (2) Gradient-free methods such as PoisonedRAG-B~\cite{zou2025poisonedrag} append content extracted from the target sample to the attack segment; however, in the cross-modal setting, the target is an image, and an unoptimized caption derived from it cannot reliably bridge the modality gap.

To address these issues, the optimization should both \textit{explore} the
embedding space and \textit{exploit} high-reward regions within it. Motivated
by recent approaches that leverage RL for prompt
optimization~\cite{deng2022rlprompt,yun2025learning}, we propose
\textbf{Reward-Guided Policy Optimization (RGPO)}, which models the retrieval
segment as a factorized stochastic policy
$\pi(T_{\text{ret}} \mid T_{\text{ext}}) = \prod_{l=1}^{L} \pi_{l}(T_{\text{ret}}[l] \mid T_{\text{ext}})$,
where $\pi_{l}$ is the token distribution at position $l$. At iteration $t$,
RGPO samples $B$ candidates $\{T_{\text{ret}}^{(i)}\}_{i=1}^B \sim \pi_{t}$,
scores them by the contrastive reward $r_i = f(T_{\text{ret}}^{(i)}; I, T_{\text{ext}})$,
and assigns each a sharpened, elite-gated weight
$w_{t,i} = \exp(r_i/\tau)\,\mathbf{1}\{r_i \ge \gamma_t\}$, where $\gamma_t$ is
the $(1-\rho)$-quantile of $\{r_i\}$ and $\tau>0$ is a temperature. For each
position $l$, the marginal is then updated toward the reweighted empirical
token frequencies:
\[
\pi_{t+1,l}(v \mid T_{\text{ext}}) = (1-\beta)\, \pi_{t,l}(v \mid T_{\text{ext}}) + \beta \sum_{i=1}^B \tilde{w}_{t,i}\, \mathbf{1}\{T_{\text{ret}}^{(i)}[l] = v\},
\]
with $v \in \mathcal{V}$ a vocabulary token, $\tilde{w}_{t,i} = w_{t,i}/\sum_j w_{t,j}$, and $\beta \in (0,1]$ controlling the adaptation speed. Intuitively, RGPO shifts probability mass toward tokens that appear frequently among high-reward candidates, while the factorized policy preserves enough exploration to escape local optima. 

\noindent
\textbf{Theoretical guarantee.} Write $f(x)=f(x;I,T_{\text{ext}})$ for the contrastive reward and assume that it decomposes additively across positions $f(x)=\sum_{l=1}^{L} f_l(x[l])$ for $x \in \mathcal V^L$ (we discuss in Appendix~\ref{app:rgpo-proof} why a first-order surrogate is locally additive). The following theorem formalizes the token-amplification effect underlying RGPO.
\begin{theorem}[RGPO $(\varepsilon,\delta)$-guarantee]\label{thm:rgpo}
Fix any $\varepsilon>0$. If the number of RGPO iterations satisfies $n = \Omega(\log |\mathcal V|)$ and the number of candidates at each iteration satisfies $B = \Omega(\log(1/\delta))$, then $\widehat T_{\mathrm{ret}}$ from the final policy satisfies
\[
f(\widehat T_{\mathrm{ret}})
\ge
\max_{x\in\mathcal V^L} f(x)-\varepsilon
\]
with probability at least $1-\delta$. See proof in Appendix~\ref{app:rgpo-proof}. 
\end{theorem}


\subsection{Extraction Segment Design}
\label{sec:ext}


The retrieval segment determines whether the test image enters the generator's context. The remaining question is how to design the extraction segment \(T_{\mathrm{ext}}\) and the scoring rule so that, together, they elicit discriminative MIA signals: low scores for non-members and high scores for members. We therefore consider a query $T = T_{\mathrm{ext}} \oplus T_{\mathrm{ret}}$ and analyze the MIA score induced by a \emph{general} scoring rule.


Let \(\mathcal X\) denote the set of retrieved images, and let $Y \sim p(y \mid \mathcal X, T)$ be the generator output. We define a measurable score function $s:\mathcal I \times \mathcal Y \to \mathbb R$, where the first argument is the test image and the second is the generated output. The single-query MIA score is the random variable $S = s(I,Y)$. Let \(Z\in\{0,1\}\) indicate whether the test image \(I\) belongs to the target database. For \(z\in\{0,1\}\), define the conditional score distribution under membership status \(z\) as
\[
F_z(a \mid I,T)
=
\Pr\!\left(S \le a \mid Z=z, I,T\right).
\]
We further define the retrieval hit event $H=\{I\in \mathcal X\}$, i.e., the event that the test image is retrieved. Let $\rho_T(I)=\Pr(H=1\mid Z=1,I,T)$ denote the retrieval probability induced by the query \(T\). We also define the hit and miss score distributions as
\[
F_{\mathrm{hit}}(a \mid I,T)
=
\Pr\!\left(S\le a \mid H=1,I,T\right),
\qquad
F_{\mathrm{miss}}(a \mid I,T)
=
\Pr\!\left(S\le a \mid H=0,I,T\right).
\]
Since a non-member image cannot be retrieved from the target database, we have \(H=0\) whenever \(Z=0\). Hence, the member and non-member score distributions satisfy the mixture identity
\[
F_1(a \mid I,T)
=
\rho_T(I) F_{\mathrm{hit}}(a \mid I,T)
+
\bigl(1-\rho_T(I)\bigr)F_{\mathrm{miss}}(a \mid I,T),
\quad
F_0(a \mid I,T)
=
F_{\mathrm{miss}}(a \mid I,T).
\]

This identity separates the two roles of our attack: the retrieval segment optimized by RGPO increases the retrieval probability \(\rho_T(I)\), while the extraction segment and the scoring function determine the separation between \(F_{\mathrm{hit}}\) and \(F_{\mathrm{miss}}\). Thus, a successful attack requires both a large retrieval probability and a scoring rule whose hit/miss gap is large. Equivalently, the extraction segment should make outputs in the hit case systematically more similar to the test image than outputs in the miss case.



Guided by this principle, we instantiate the extraction segment as a prompt that asks the generator to describe the retrieved reference images. We then define the MIA score as the similarity between the test image \(I\) and the generated output \(Y\). Because T2I and Q\&A systems expose different interaction patterns, we employ task-specific adaptations of this extraction-and-scoring strategy, described below.


\textbf{(a) T2I.} Since T2I generators~\citep{podell2023sdxl} lack instruction tuning, prompts such as \textit{``please describe the retrieved images''} are largely ignored (low text attention) and easily flagged as anomalous requests (as illustrated in Appendix~\ref{app:detection_details}). To address this, we use the caption of the test image as the extraction segment, leveraging the copy-like behavior of generators under aligned inputs. If $I$ is successfully retrieved as a reference (i.e., a member), the T2I generator faithfully reproduces its content. Even when the retrieved image is not $I$, the generator still tends to reproduce salient content from the reference, as it shares similar semantics with the caption. Together, these effects make caption-based extraction an effective realization of our design principle. As illustrated in Appendix~\ref{app:qual}, members trigger near-verbatim reconstruction through the copy-like behavior, while non-members yield only semantically aligned but visually divergent outputs. \textbf{(b) Q\&A.} A direct command such as \textit{``describe each retrieved image''} causes the VLM to enumerate all retrieved images in a single response, introducing substantial noise into the similarity computation. To address this, we split the output into per-image descriptions and define the MIA score as the maximum similarity between $I$ and any individual description. As shown in Table~\ref{tab:scoring}, we further observe that computing similarity within the same modality (i.e., T2T or I2I) is essential for achieving strong performance.

\subsection{The Overall Pipeline and Multi-query Aggregation}\label{sec:pipeline}

Our single-query attack proceeds in four steps: (1) construct an extraction segment $T_{\text{ext}}$ following Sec.~\ref{sec:ext}; (2) construct a retrieval segment $T_{\text{ret}}$ following Sec.~\ref{sec:ret}; (3) feed the composite query $T = T_{\text{ext}} \oplus T_{\text{ret}}$ into the target system and obtain its output; and (4) compute the MIA score: image-to-image similarity between the test image and the output in the T2I setting, or the maximum text-to-text similarity between the test image description and each output fragment in the Q\&A setting.


\textbf{Multi-query aggregation.} Some queries may fail to elicit a meaningful MIA signal due to errors in optimizing the retrieval segment or approximating the appropriate extraction segment. To improve reliability, we issue $M \times N$ queries by pairing each of $M$ extraction segments with $N$ optimized retrieval segments. We then apply K-means clustering to the resulting scores, retain only the high-scoring cluster, and aggregate the retained scores to produce the final MIA decision.


\vspace{-2mm}
\section{Experiments}
We aim to answer the following research questions: (1) How does {\name} perform compared to adapted baselines under both T2I and Q\&A settings? (Sec.~\ref{sec:exp_main}) (2) Is {\name} robust under various settings? (Sec.~\ref{sec:exp_abl}) (3) How do our attack designs contribute to {\name}? (Sec.~\ref{sec:exp_abl})

\subsection{Experimental Setup}\label{sec:experiments}
\textbf{Image database and evaluation datasets.} 
For both T2I and Q\&A tasks, we construct large-scale image databases following prior work~\cite{lyu2025realrag,chen2024mllm}. \textbf{(1) T2I tasks.} We construct a database of 259{,}057 images spanning six datasets: MSCOCO~\cite{lin2014microsoft}, ImageNet-100 subset~\cite{russakovsky2015imagenet}, WikiArt~\cite{phillips2011wiki}, Stanford Cars~\cite{krause20133d}, Stanford Dogs~\cite{dataset2011novel}, and CelebA-HQ~\cite{karras2017progressive}. We use each dataset's training split to construct the database, and sample member and non-member images from the training and test splits. Due to computational constraints, we randomly select three of the six datasets for attack evaluation. Details can be found in Appendix~\ref{app:dataset}. For utility evaluation, we follow the standard protocol~\cite{lyu2025realrag} as illustrated in Appendix~\ref{app:dataset}. \textbf{(2) Q\&A tasks.} Following~\cite{chen2024mllm}, we use MMQA~\cite{talmor2021multimodalqa} and the MSCOCO training split as the database, containing 58{,}075 and 118{,}287 images, respectively. Since MMQA does not provide a training/test split, we randomly hold out 1{,}000 images as non-members and treat the rest as members. For utility evaluation, we follow the same setup as AQUA~\cite{chen2025safeguarding}.

\textbf{Models.} For T2I tasks, we employ four generators: SDXL~\cite{podell2023sdxl} plus IP-Adapter~\cite{ye2023ip} (default), SD1.5~\cite{rombach2022high} plus IP-Adapter, Kandinsky v2.2~\cite{razzhigaev2023kandinsky}, and SDXL plus Conceptrol~\cite{he2025conceptrol}, paired with CLIP-ViT-H/14~\cite{radford2021learning} as the retriever. For Q\&A tasks, we employ Qwen2.5-VL-7B-Instruct~\cite{wang2024qwen2} (default) and LLaVA-1.6-Mistral-7B~\cite{liu2023visual} as generators, with CLIP-ViT-L/14-336 as the retriever. The setup follows standard practice in the {\rag} literature~\cite{lyu2025realrag,chen2024mllm}.

\textbf{Metrics.} We report AUROC, ACC, TPR@5\%FPR, and Retrieval Success Rate (RSR) for attack evaluation. Details of the attack and utility evaluation metrics are provided in Appendix~\ref{app:metrics}.

\textbf{Baselines.} We compare {\name} against four baselines. \textit{Na\"ive} directly uses the caption of the test image as the query, with no retrieval segment. \textit{PoisonedRAG-B$^{*}$}, \textit{PoisonedRAG$^{*}$} and \textit{BadRAG$^{*}$} are adapted baselines that retain our extraction segment but optimize the retrieval segment using techniques from existing RAG poisoning attacks~\cite{zou2025poisonedrag,xue2024badrag} (See Appendix~\ref{app:baselines}). We do not include previous text-RAG MIAs, as their Q\&A-based pipelines do not transfer to {\rag}. We also exclude jailbreaking methods like GCG~\cite{zou2023universal}, as they are less effective than PoisonedRAG in this setting~\cite{zou2025poisonedrag}.

\begin{table*}[t]
\centering
\caption{Attack results of {\name} and adapted baselines$^*$ on {\rag} for T2I tasks. For adapted baselines, we use the same extraction segments and optimize retrieval segments using their respective methods. The system uses CLIP ViT-H/14 for retrieval and SDXL+IP-Adapter for generation. TPR$_{5\%}$ denotes TPR@5\%FPR. Detailed TPR-FPR curves are provided in Appendix Figure~\ref{fig:exp_t2i_curves}.}
\label{tab:t2i_attack_results}
\small
\setlength{\tabcolsep}{3pt}
\renewcommand{\arraystretch}{0.9}

\begin{tabular}{lcccc cccc cccc}
\toprule
\textbf{Attack Method} 
& \multicolumn{4}{c}{\textbf{MSCOCO}}
& \multicolumn{4}{c}{\textbf{WikiArt}} 
& \multicolumn{4}{c}{\textbf{Stanford Dogs}} \\
\cmidrule(lr){2-5} 
\cmidrule(lr){6-9} 
\cmidrule(lr){10-13}

& AUC & ACC & {\tprfpr} & RSR
& AUC & ACC & {\tprfpr} & RSR
& AUC & ACC & {\tprfpr} & RSR \\
\midrule

Na\"ive
& 0.72 & 0.70 & 0.37 & 0.53 & 0.63 & 0.63 & 0.31 & 0.33 & 0.66 & 0.67 & 0.29 & 0.40 \\
\cmidrule(lr){1-13}

PoisonedRAG-B$^*$ 
&0.71 & 0.71 & 0.26 & 0.55 & 0.68 & 0.66 & 0.33 & 0.42 & 0.70 & 0.68 & 0.38 & 0.48 \\

PoisonedRAG$^*$ 
& 0.79 & 0.77 & 0.45 & 0.66 & 0.69 & 0.70 & 0.40 & 0.53 & 0.72 & 0.74 & 0.40 & 0.52 \\

BadRAG$^*$ 
& 0.80 & 0.78 & 0.46 & 0.70 & 0.66 & 0.67 & 0.30 & 0.48 & 0.71 & 0.71 & 0.33 & 0.54 \\

{\name} 
& \textbf{0.91} & \textbf{0.85} & \textbf{0.55} & \textbf{0.96} 
& \textbf{0.87} & \textbf{0.82} & \textbf{0.49} & \textbf{0.93} 
& \textbf{0.90} & \textbf{0.84} & \textbf{0.65} & \textbf{0.90} \\
\bottomrule
\end{tabular}
\end{table*}
\begin{figure*}[t]
    \centering
    \includegraphics[width=0.99\textwidth]{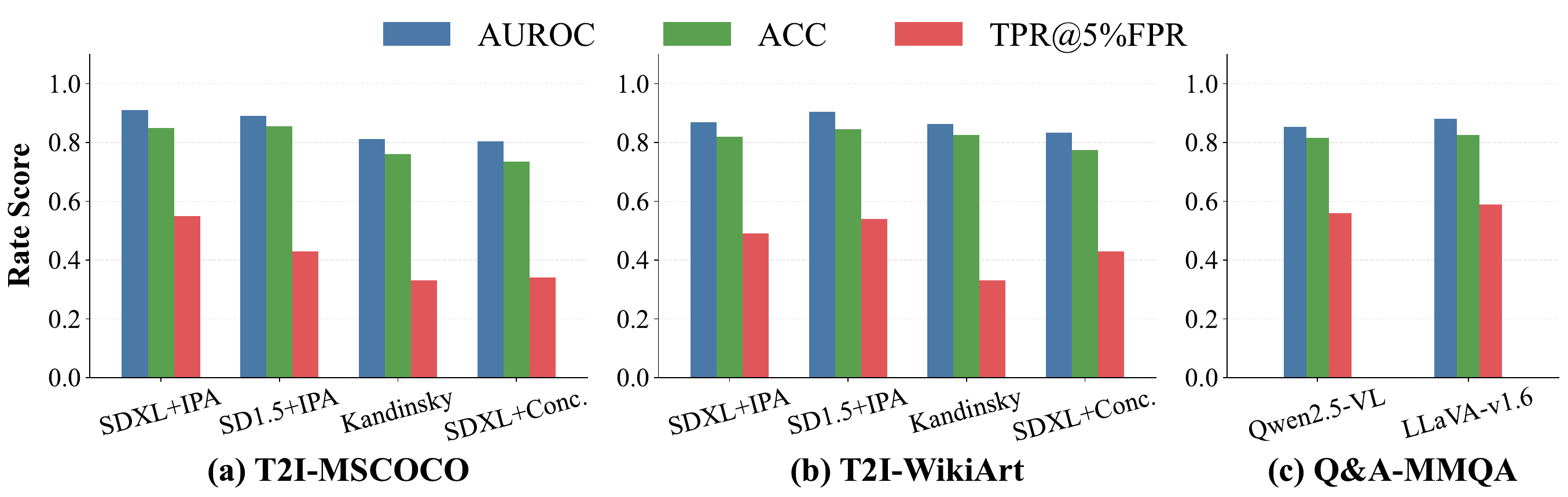}
    \caption{{\name} remains effective across different generators for both T2I and Q\&A tasks.}
    \label{fig:exp_generator}
    \vspace{-3mm}
\end{figure*}

\textbf{Evaluation setup.} 
Unlike prior work, our image database contains samples drawn from multiple sources (e.g., MSCOCO, WikiArt). In practice, an auditor is typically interested in a specific data domain (e.g., artwork) and collects calibration samples from that domain to set the threshold and infer membership. We therefore report per-dataset results, which better simulate this realistic auditing scenario, and defer the combined attack setting to Appendix Table~\ref{tab:full_mia}.

\textit{(1) Attack settings.} The attacker is allowed $M \times N$ queries per test image, where $M$ is the number of extraction segments and $N$ is the number of retrieval segments optimized per extraction segment; we set $M = N = 2$ by default. For the extraction segment, we use Qwen-VL to generate and rank candidate captions in the T2I setting, and manually design structured prompts in the Q\&A setting. For the retrieval segment, we set the token length to 8 and sample a small shadow set of 32 images (disjoint from the database). By default, the attacker has access to $E_{\text{tgt}}$; under the restrictive setting, we instead assume access to shadow embedding models, which are smaller and trained with different data distributions~\cite{xu2023demystifying,sun2023eva}. The scoring model is the target (or a selected shadow) embedding model for T2I and a text-embedding model E5~\cite{wang2022text} for Q\&A. Due to the high experimental cost ($\sim$1\,min/sample), we use a 200 member/non-member set per subdataset and verify the stability in Appendix Table~\ref{tab:stability}. Remaining hyperparameters (optimization iterations, RGPO temperature, etc.) are deferred to Appendix~\ref{app:hyper}. \textit{(2) {\rag} settings.} The retriever uses $K = 1$ for T2I and $K = 2$ for Q\&A by default, following the literature~\cite{ye2023ip,chen2025safeguarding}; this small retrieval budget poses a substantial challenge for our attack. Other generation settings follow each generator's default setup (Appendix~\ref{app:details2}).

\begin{figure*}[t]
    \centering
    \includegraphics[width=0.99\textwidth]{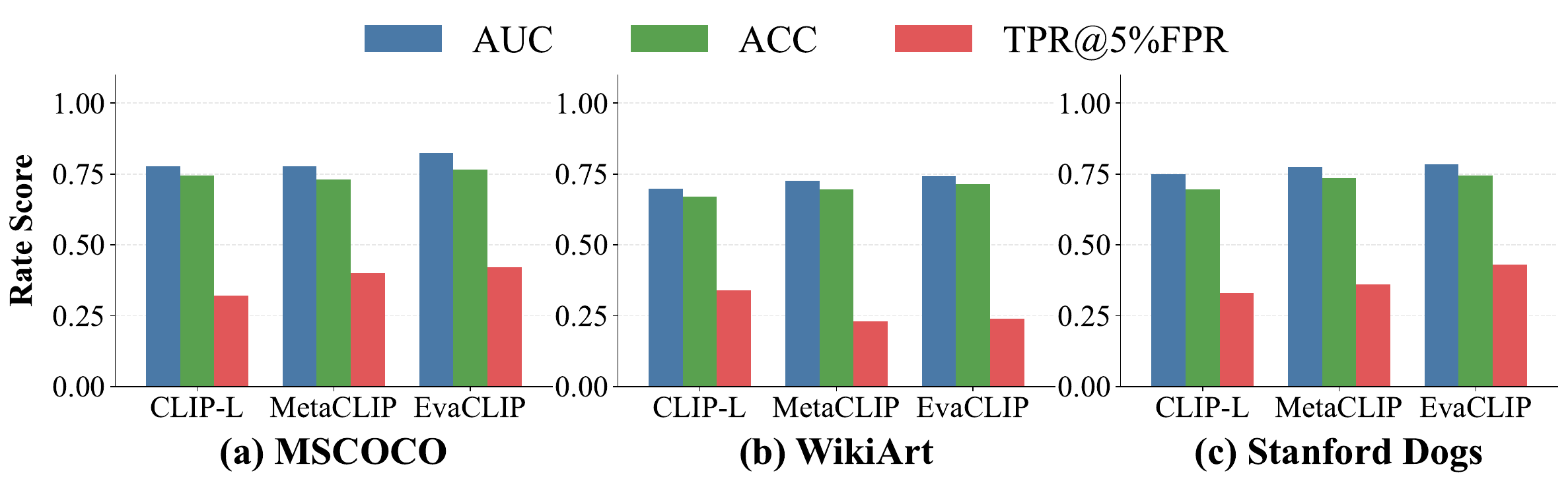}
    \caption{{\name} remains effective under strict black-box settings. We use a weaker shadow embedding model (e.g., with different architectures or training distributions) to optimize the retrieval segments and compute the MIA scores. The target system uses CLIP ViT-H/14 for retrieval.}
    \label{fig:exp_black}
\end{figure*}

\begin{table*}[t]
\centering
\label{tab:ablations}
\small
\setlength{\tabcolsep}{3.pt}
\renewcommand{\arraystretch}{0.8}
\begin{minipage}{0.48\textwidth}
\centering
\captionof{table}{Fine-grained captions for the extraction segment elicit stronger copy-like behavior and stronger attacks than generic captions (Gen.).}
\label{tab:exp_ext_main}
\begin{tabular}{lccc ccc}
\toprule
\multirow{2}{*}{\shortstack{\textbf{Type}}}
& \multicolumn{3}{c}{\textbf{MSCOCO}}
& \multicolumn{3}{c}{\textbf{WikiArt}} \\
\cmidrule(lr){2-4}
\cmidrule(lr){5-7}
& AUC & ACC & {\tprfpr}
& AUC & ACC & {\tprfpr} \\
\midrule
Gen. & 0.71 & 0.70 & 0.38 & 0.75 & 0.72 & 0.39 \\
Fine. & \textbf{0.91} & \textbf{0.85} & \textbf{0.55} & \textbf{0.87} & \textbf{0.82} & \textbf{0.49} \\
\bottomrule
\end{tabular}
\end{minipage}
\hspace{0.01\textwidth}
\begin{minipage}{0.48\textwidth}
\centering
\captionof{table}{{\name} achieves stronger attacks when scoring within a single modality (Single) than across modalities (Cross).}
\label{tab:scoring}
\begin{tabular}{c|cccccc}
\toprule
\textbf{Type} & \multicolumn{3}{c}{\textbf{T2I-MSCOCO}}
& \multicolumn{3}{c}{\textbf{Q\&A-MMQA}} \\
\cmidrule(lr){2-4} \cmidrule(lr){5-7}  
& \textbf{AUC}
& \textbf{ACC} & \textbf{\tprfpr} & \textbf{AUC}
& \textbf{ACC} & \textbf{\tprfpr} \\
\midrule
Cross & 0.60 & 0.59 & 0.14 & 0.82 & 0.79 & 0.37 \\
Single & \textbf{0.91} & \textbf{0.85} & \textbf{0.55} & \textbf{0.85} & \textbf{0.82} & \textbf{0.56} \\
\bottomrule
\end{tabular}
\end{minipage}
\end{table*}

\subsection{Main Results}
\label{sec:exp_main}

In this part, we discuss how different MIAs perform against {\rag}. We further validate that {\rag} substantially improves utility on both T2I and Q\&A tasks, and defer this analysis to Appendix~\ref{app:experiments}.

\textbf{{\name} outperforms baseline attacks on {\rag} systems for T2I and Q\&A tasks.} We present the main results of {\name} and adapted baselines on {\rag} for T2I tasks in Table~\ref{tab:t2i_attack_results} and for Q\&A tasks in Appendix Table~\ref{tab:t2t_attack_results}. All methods share the same extraction segment and differ only in retrieval segment optimization. (1) Our attack achieves state-of-the-art performance on both T2I and Q\&A tasks, outperforming existing attacks by a large margin (e.g., over 10\% AUROC on T2I tasks). Existing attacks suffer from low target retrieval rates because they easily get trapped in local minima within the cross-modal embedding space. (2) Our attack is effective in the low false positive region (also illustrated in Figure~\ref{fig:exp_t2i_curves} and~\ref{fig:exp_t2t_curves}), achieving around 50\% TPR@5\%FPR, which is commonly interpreted as a strong attack in the literature~\citep{carlini2022membership}. (3) In the T2I setting, WikiArt shows lower performance than other datasets because diffusion models exhibit weaker copy-like behavior on artwork references, as stylized images are less aligned with the text. (4) In the Q\&A setting, our attack is more effective on MMQA due to the uniqueness of its images (e.g., film posters). (5) Existing baselines perform poorly in the Q\&A setting because when the extraction segment (``Describe each retrieved...'') is unrelated to the test image, the initial query embedding lies far from the target, making gradient-based optimization challenging-a regime where RGPO's reward-guided updates succeed.


\textbf{{\name} remains effective across diverse generators for both T2I and Q\&A tasks.} For T2I, beyond the default SDXL+IPA, we evaluate generators varying along two axes—diffusion backbone and conditioning mechanism—including SD1.5+IPA (different backbone), Kandinsky (different family, adapter-free), and SDXL+Conceptrol (alternative conditioning). As shown in Figure~\ref{fig:exp_generator}(a,b), {\name} maintains high AUROC (>0.80) across all variants. Conceptrol yields the weakest attack due to its more creative generation capability, yet it still exceeds 0.30 TPR@5\%FPR. For Q\&A, Qwen2.5-VL and LLaVA-v1.6 produce nearly identical results (Figure~\ref{fig:exp_generator}(c)), as both VLMs faithfully describe image content. Additional results are provided in the Appendix Figure~\ref{fig:exp_generator2}. Overall, these results confirm that {\name} is robust to the choice of underlying generators.


\begin{figure*}[t]
    \centering
    \includegraphics[width=0.88\textwidth]{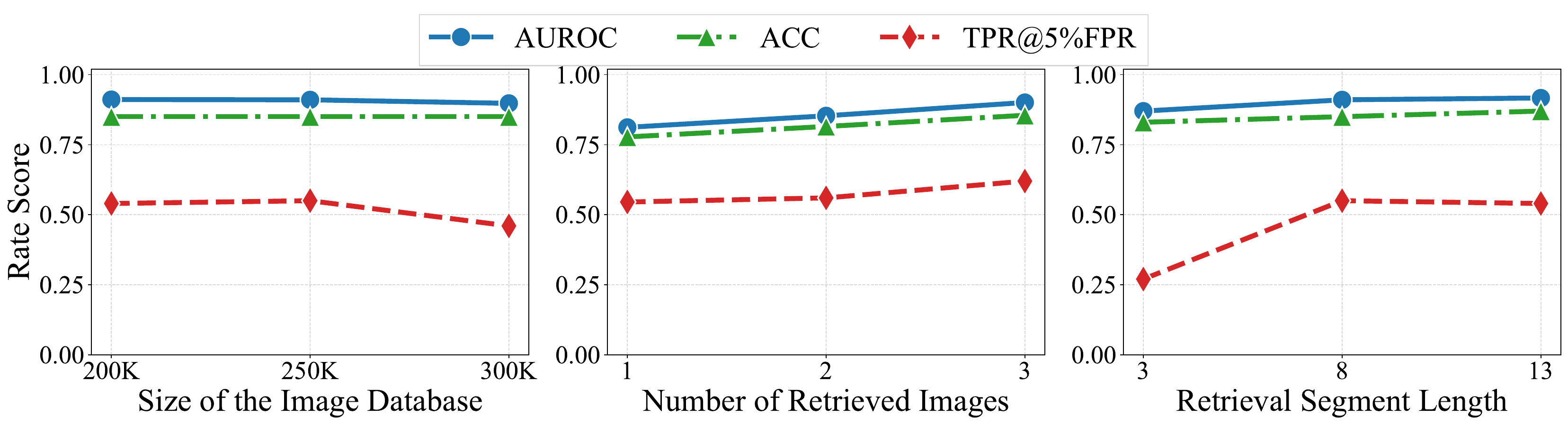}
    \caption{Impact of (a) the size of the image database, and (b) the number of retrieved images, and (c) the retrieval segment length on attack results. {\name} remains robust across all these settings.}
    \label{fig:exp_rag_abl}
    \vspace{-2mm}
\end{figure*}

\subsection{Ablation Study}
\label{sec:exp_abl}
Unless otherwise mentioned, we conduct ablation studies on {\rag} for T2I tasks.

\textbf{{\name} remains effective in the restrictive attack setting.} Under restrictive constraints, where the attacker has no access to the embedding model used by the target system, we construct retrieval segments using a weaker shadow embedding model. Despite this mismatch, Figure~\ref{fig:exp_black} shows that {\name} still achieves an AUROC of around 0.75, which is generally considered strong for MIA. The observed performance drop stems not only from weaker retrieval segment construction but also from the weaker scoring model used in this setup; as shown in Appendix Table~\ref{tab:scoring_model}, stronger scoring models consistently yield better results by capturing more fine-grained image semantics. Interestingly, EvaCLIP transfers best on MSCOCO and Dogs but degrades on WikiArt, likely due to lower embedding similarity with the target (ViT-H/14) on artwork images.

\textbf{Our tailored attack designs lead to stronger attack results.} We further investigate the impact of our attack designs for the \textit{extraction segment}, \textit{retrieval segment}, and \textit{scoring mechanism}. For the extraction segment, as shown in Table~\ref{tab:exp_ext_main} and Appendix Table~\ref{tab:exp_extraction}, fine-grained captions generated by BLIP and Qwen-VL significantly improve the attack results, yielding up to $\sim$10\% gains in AUROC. We attribute this to the fact that high-quality captions more effectively elicit copy-like behavior from diffusion models and provide a better starting point for retrieval segment construction. For the retrieval segment, we have demonstrated the superiority of the proposed RGPO mechanism in Table~\ref{tab:t2i_attack_results}. In Appendix Table~\ref{tab:exp_shadow}, we further evaluate the impact of incorporating a small shadow set for contrastive reward. Under the restrictive setting, we show that negative samples consistently improve attack performance by modeling a more favorable optimization landscape that generalizes. Beyond these two components, we also examine different strategies for computing the similarity score. In both T2I and Q\&A settings, similarity can be measured between the output and the image (image modality) or between the output and image descriptions (text modality). As shown in Table~\ref{tab:scoring}, we observe that computing similarity within the same modality yields large improvements, especially for TPR$_{5\%}$; accordingly, we extract image details from the test image in the Q\&A setting. 

\textbf{{\name} remains effective under different {\rag} setups.} In Figure~\ref{fig:exp_rag_abl}, we investigate the impact of the size of the image database, the number of retrieved images, and the length of the retrieval segment on attack results. In (a), we enlarge the image database by incorporating additional images from ImageNet and observe that the database size has only a minor effect on attack performance. This is because RGPO can effectively retrieve the test image whenever it lies in the database. We further observe that a larger retrieval budget slightly improves the attack in the Q\&A setting, as K=3 increases the likelihood of the members being retrieved. Here, we experiment with Q\&A setting due to architectural constraints in T2I generators. Finally, {\name} saturates even when the retrieval segment length is small (8 tokens), which makes our attack stealthier and less likely to be detected.

\textbf{Ablation on Multi-Query Attack Strategies.} We evaluate the impact of aggregation strategy and query number for multi-query attacks in Appendix Figures~\ref{fig:exp_aggregation} and~\ref{fig:exp_queries}. We find that the K-means-based strategy balances precision and coverage, yielding more stable attack performance across settings. We also observe that increasing the number of queries improves attack results, though notably, even a single-query attack achieves competitive performance. Detailed analysis is provided in Appendix~\ref{app:experiments}.

\vspace{-1mm}
\section{Conclusion}
We introduce \textit{\name}, the first membership inference attack (MIA) against image-based RAG (IRAG) systems. To overcome fundamental limitations of prior text-RAG MIAs, \textit{\name} decomposes each attack query into a retrieval segment and an extraction segment. We propose Reward-Guided Policy Optimization to enhance cross-modal retrieval, and design tailored prompting strategies paired with a scoring rule to elicit discriminative MIA signals. Extensive experiments show that \textit{\name} achieves over 80\% AUROC with only four queries and remains effective under restrictive attack settings. These results establish {\name} as a practical auditing tool.

\textbf{Limitations and Future Work.}
This work focuses on establishing the feasibility of MIAs against IRAG. While our attack achieves strong performance across multiple settings, we evaluate robustness only against text/image attention ratio-based detection (see Appendix~\ref{app:detection_details}); more sophisticated defenses, such as differential privacy~\cite{koga2024privacy} and output perturbation, remain unexplored. In addition, we focus on text-based retrieval and do not evaluate multimodal inputs. However, as discussed in Appendix~\ref{app:discussion}, our attack naturally extends to this setting. We leave these directions to future work.


\bibliographystyle{plainnat}
\bibliography{main}


\appendix
\newpage
\section{Proofs}\label{app:theory}

\subsection{Proof of Theorem~\ref{thm:rgpo}}
\label{app:rgpo-proof}

We prove the theorem for $\beta = 1$. $\beta < 1$ introduces additional approximation error but does not change the qualitative amplification mechanism.

For each position $l\in\{1,\ldots,L\}$, define
\[
\mathcal G_l := \Big\{v\in\mathcal V: f_l(v)\ge \max_{v\in\mathcal V} f_l(v) - \frac{\varepsilon}{L}\Big\}.
\]
If $\mathcal G_l=\mathcal V$, then every token at position $l$ is already $\varepsilon/L$-good and there is nothing to prove for that position. Otherwise define the good--bad margin $\Delta_l := \min_{v\in\mathcal G_l} f_l(v) - \max_{v\notin\mathcal G_l} f_l(v) > 0$ and $\Delta:=\min_{l} \Delta_l >0$.

We note that any sequence whose token at each position lies in the corresponding good set is $\varepsilon$-optimal. To see that, for any $x\in\mathcal V^L$ with $x[l]\in\mathcal G_l$ for all $l$, we have
\begin{align*}
\max_{z\in\mathcal V^L} f(z) - f(x) =
\sum_{l=1}^L \left(\max_{v\in\mathcal V} f_l(v)-f_l(x[l])\right) \le
\sum_{l=1}^L \frac{\varepsilon}{L} = \varepsilon.
\end{align*}
Therefore, it suffices to show that the modal token of the final policy belongs to $\mathcal G_l$ at every position.

To see the token-amplification effect, we first analyze the one-step amplification. Fix an iteration $t$ and a position $l$. Let the current product policy be $\pi_t=\prod_{j=1}^L \pi_{t,j}$, and let $X\sim\pi_t$. Define the importance weight $w_t(x):= \exp(f(x)/\tau) \mathbf{1}\{f(x) \ge \gamma_t\}$. For each token $v\in\mathcal V$, let $m_{t,l}(v) := \mathbb E[w_t(X)\mid X[l]=v,\pi_t]$. Since the reward is additive, if we write $S_{t,l} := \sum_{j\neq l} f_j(X[j]),$ then we have
\[
m_{t,l}(v) = \exp\left(\frac{f_l(v)}{\tau}\right)
\mathbb E\left[
\exp\left(\frac{S_{t,l}}{\tau}\right)
\mathbf 1\{S_{t,l}\ge \gamma_t-f_l(v)\}
\middle|\, \pi_t
\right].
\]
For any $v\in\mathcal G_l$ and $u\notin\mathcal G_l$, since $f_l(v)-f_l(u)\ge \Delta_l\ge\Delta$, we have
\begin{align*}
m_{t,l}(v) &\ge \exp\left(\frac{f_l(v)-f_l(u)}{\tau}\right)
\exp\left(\frac{f_l(u)}{\tau}\right)
\mathbb E\left[
\exp\left(\frac{S_{t,l}}{\tau}\right)
\mathbf 1\{S_{t,l}\ge \gamma_t^{\star}-f_l(u)\}
\middle|\, \pi_t
\right] \\
&\ge \exp\left(\frac{\Delta}{\tau}\right) m_{t,l}(u).
\end{align*}

Define the exact marginal update induced by $w_t$
\begin{align*}
p_{t+1,l}(v)= \frac{\mathbb E[w_t(X)\mathbf 1\{X[l]=v\}\mid \pi_t]}{\mathbb E[w_t(X)\mid \pi_t]} = \frac{\pi_{t,l}(v)m_{t,l}(v)}{\sum_{u\in\mathcal V} \pi_{t,l}(u)m_{t,l}(u)},
\end{align*}
and define
\begin{align*}
q_{t,l} := \sum_{v\in\mathcal G_l} \pi_{t,l}(v), \qquad \bar q_{t+1,l} := \sum_{v\in\mathcal G_l} p_{t+1,l}(v).
\end{align*}
Then we have
\begin{align*}
\bar q_{t+1,l}
&=
\frac{\sum_{v\in\mathcal G_l} \pi_{t,l}(v)m_{t,l}(v)}{\sum_{v\in\mathcal G_l} \pi_{t,l}(v)m_{t,l}(v)+\sum_{v\notin\mathcal G_l} \pi_{t,l}(v)m_{t,l}(v)} \ge \frac{\kappa q_{t,l}}{(1-q_{t,l})+\kappa q_{t,l}},
\end{align*}
where $\kappa = \exp\left(\Delta/\tau\right) > 1$. Equivalently,
\[
\frac{\bar q_{t+1,l}}{1-\bar q_{t+1,l}} \ge \kappa\frac{q_{t,l}}{1-q_{t,l}}.
\]
Therefore the update multiplies the good-token odds by at least the factor $\kappa$ at every iteration.

Given the amplification result, we can now bound the number of RGPO iterations $n$ and the number of samples at each iteration $B$. Since RGPO is initialized uniformly, for every position $l$ we have $q_{1,l} \ge 1/|\mathcal V|$. The amplification result then implies that, after $n = \Omega(\log|\mathcal V|)$ iterations, the good-token mass at each position satisfies $q_{n,l} \ge 1 - 1/(2L)$. Therefore, for a single sample $X \sim \pi_n$ from the final policy, a union bound over the $L$ positions gives $\Pr(X[l] \in \mathcal G_l \text{ for all } l) \ge 1-\sum_{l=1}^L \Pr(X[l] \notin \mathcal G_l) \ge 1 - L/(2L) = 1/2$. Whenever this event occurs, the additive structure of $f$ implies $f(X) \ge \max_{x\in\mathcal V^L} f(x)-\varepsilon$. Since each draw is $\varepsilon$-optimal with probability at least 1/2, by the independent sampling from the policy, choosing $B = \Omega(\log(1/\delta))$ ensures that, with probability at least $1-\delta$,
\[
f(\widehat T_{\mathrm{ret}})
\ge
\max_{x\in\mathcal V^L} f(x)-\varepsilon.
\]
This completes the proof.

\paragraph{On the additivity assumption.}
The cosine reward is not globally position-additive. However, for a fixed
extraction segment $T_{\text{ext}}$ and a short retrieval suffix,
a first-order Taylor expansion of
$\mathrm{sim}\bigl(E(T_{\text{ext}} \oplus T_{\text{ret}}),\, E(I)\bigr)$
around the current policy mean is position-separable:
\[
f(x) \;\approx\; c \;+\; \sum_{l=1}^{L} \langle g_l,\, \phi(x[l]) \rangle,
\]
where $g_l$ is the gradient with respect to the $l$-th token embedding and
$\phi$ is the embedding lookup. Theorem~\ref{thm:rgpo} should therefore be
read as a guarantee on this local surrogate, which RGPO effectively optimizes
through reward-ranked sampling. The empirical retrieval success rates (90\%+) in our experiments (Table~\ref{tab:t2i_attack_results}) are consistent with this approximation
being tight in the regimes we evaluate.
\newpage
\begin{figure*}[t]
    \centering
    \begin{subfigure}[t]{0.24\textwidth}
        \centering
        \includegraphics[width=\linewidth]{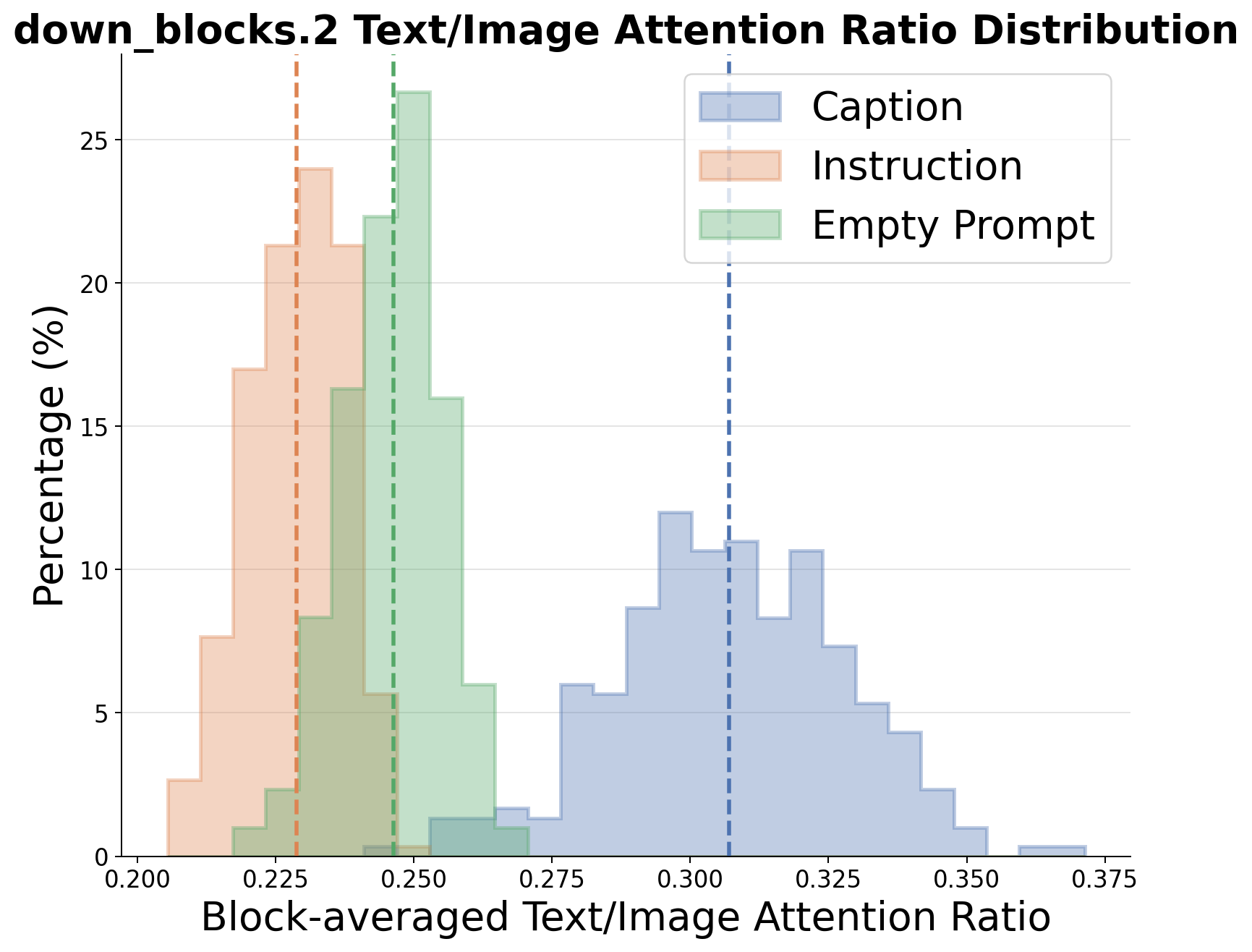}
        \caption{Diffusion Step 1}
    \end{subfigure}
    \hfill
    \begin{subfigure}[t]{0.24\textwidth}
        \centering
        \includegraphics[width=\linewidth]{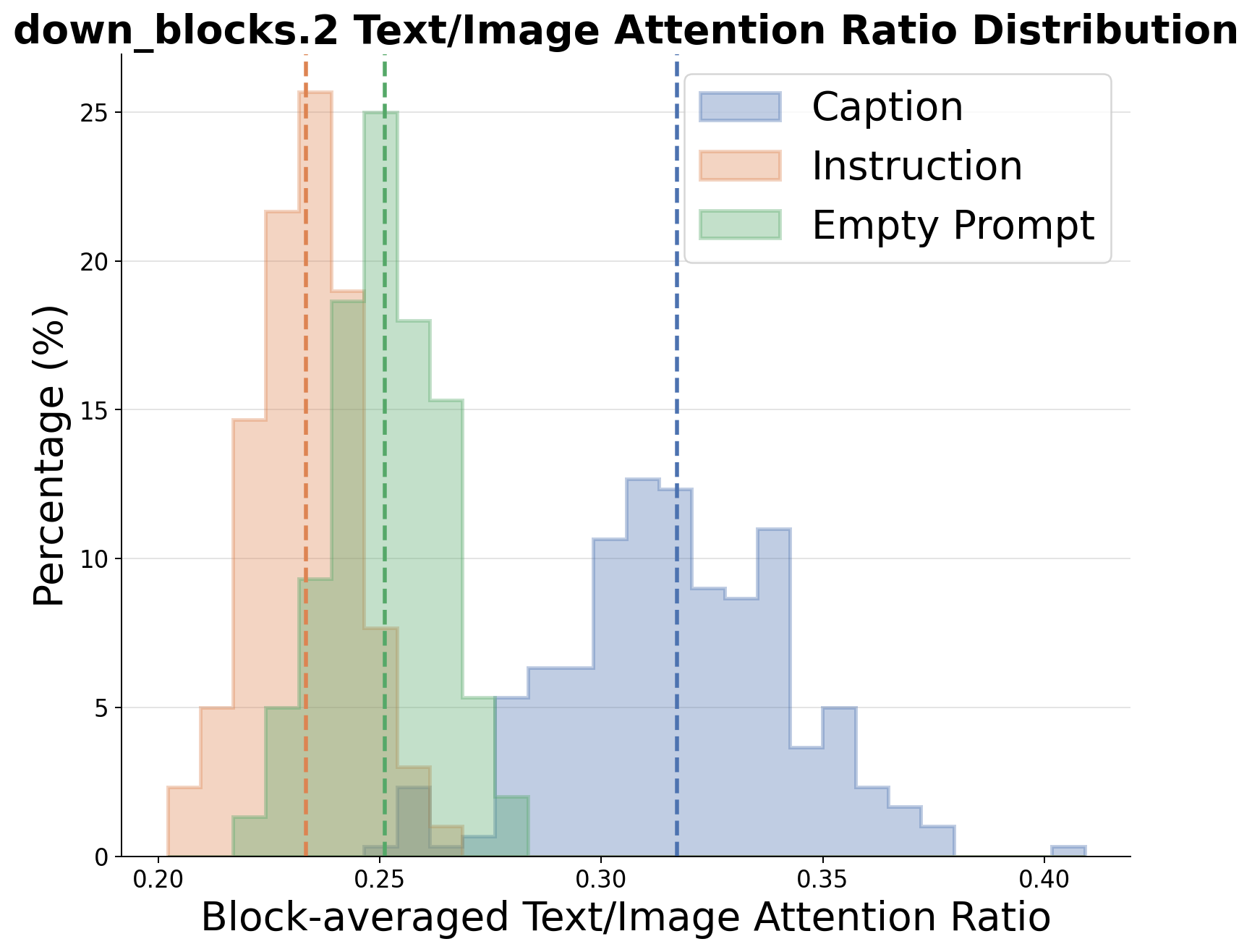}
        \caption{Diffusion Step 6}
    \end{subfigure}
    \hfill
    \begin{subfigure}[t]{0.24\textwidth}
        \centering
        \includegraphics[width=\linewidth]{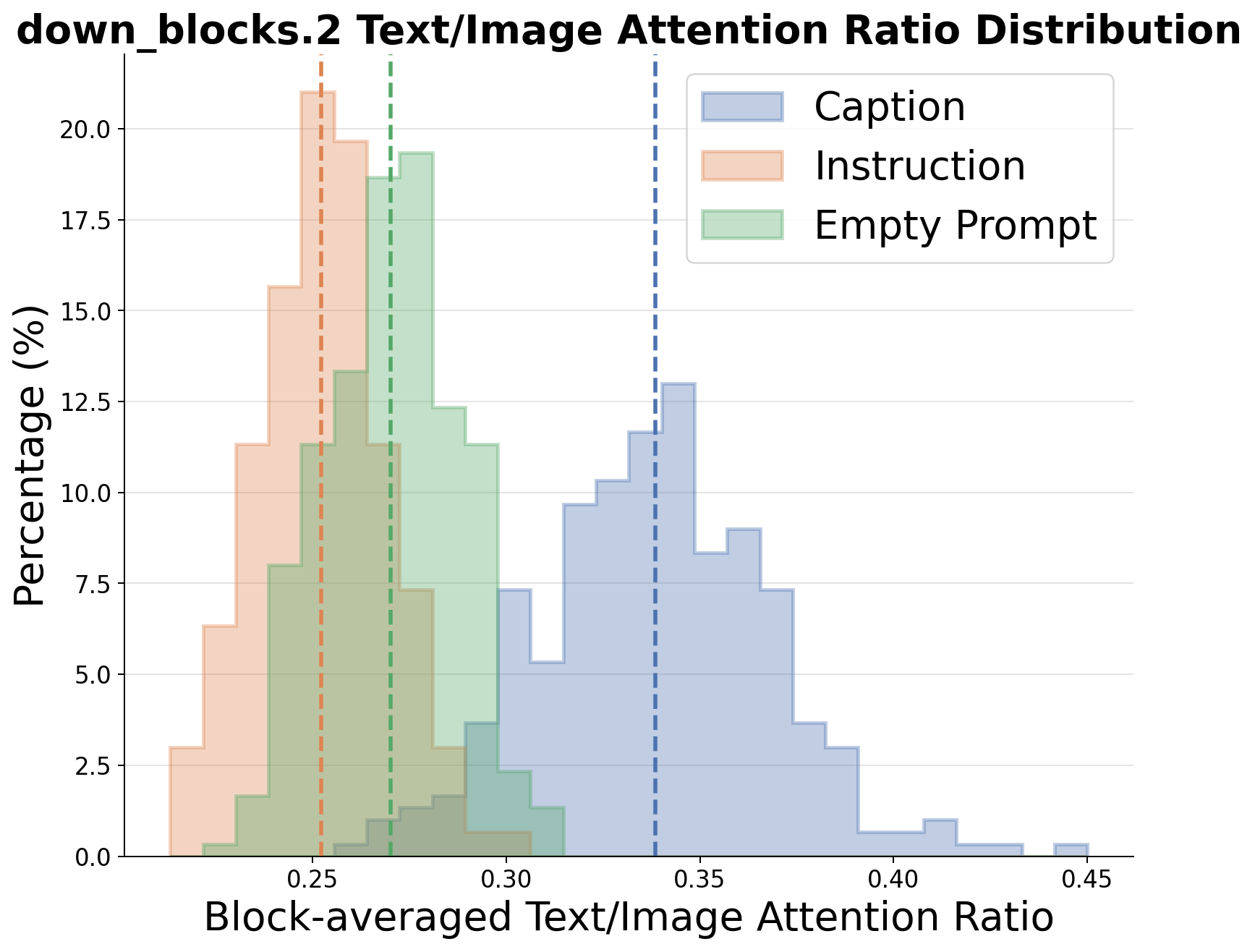}
        \caption{Diffusion Step 11}
    \end{subfigure}
    \hfill
    \begin{subfigure}[t]{0.24\textwidth}
        \centering
        \includegraphics[width=\linewidth]{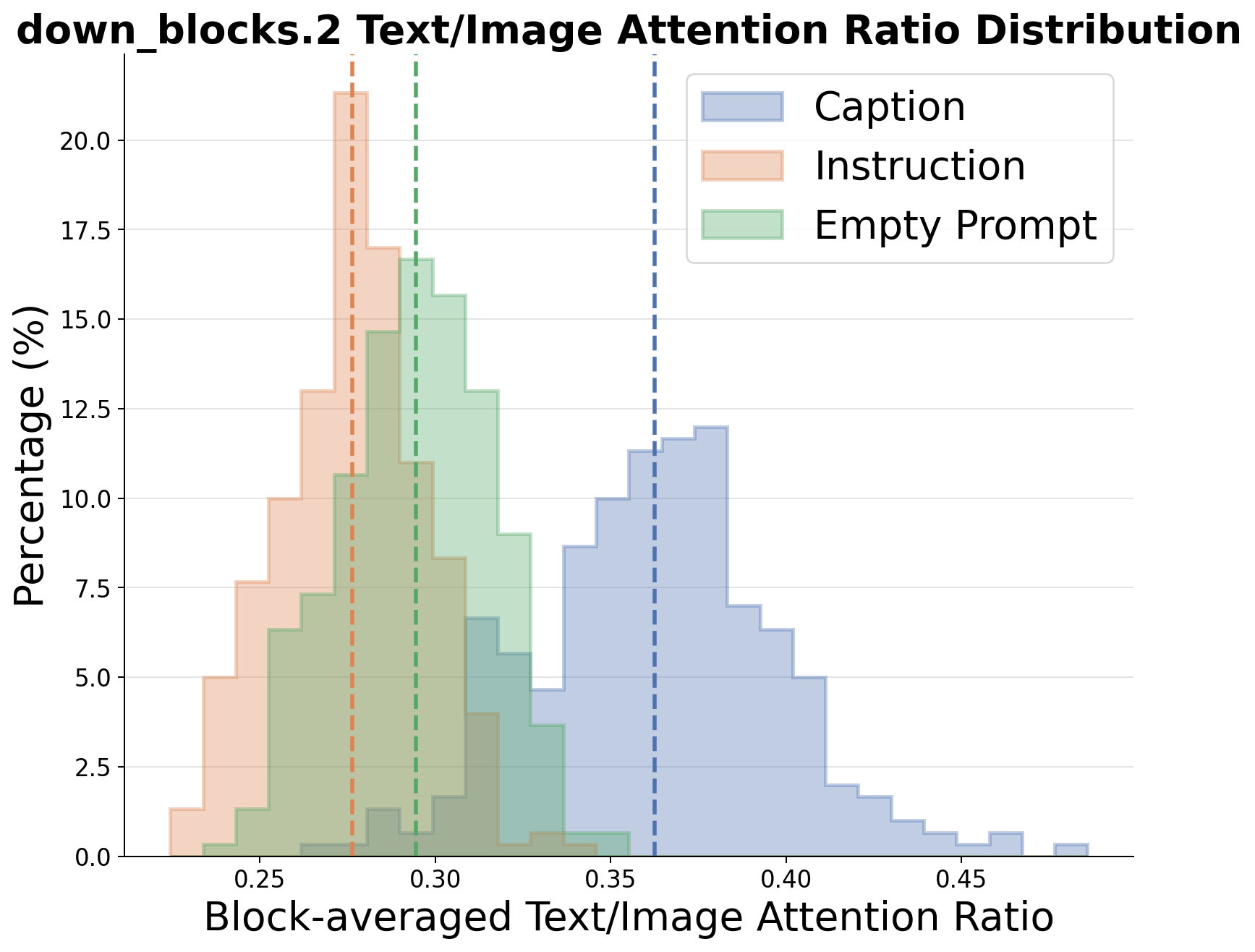}
        \caption{Diffusion Step 16}
    \end{subfigure}
    \caption{Instruction-based extraction segments produce abnormal text-image attention ratios due to the lack of semantic meaning in text prompts, which can be easily detected in early diffusion steps.}
    \label{fig:detection}
\end{figure*}

\section{Text/Image Attention Ratio based Detection}
\label{app:detection_details}
In the T2I setting, a na\"ive attacker may use an instruction prompt (e.g., \textit{``Please describe the retrieved images''}) or even an empty prompt (e.g., \textit{``''}) to coerce the T2I generator into directly outputting the retrieved image. The underlying intuition is that, in such cases, the text input is largely ignored and generation is governed primarily by the image reference, leading to direct leakage of the retrieved image. While effective in certain cases, we show that this strategy can be easily detected by a simple \textit{Text/Image Attention Ratio} mechanism.

\textbf{Detection mechanism.} Let's take SDXL+IP-Adapter as an example. The defender collects cross-attention scores from a specific U-Net block (we use \texttt{down\_blocks.2}) of the diffusion model. In the IP-Adapter architecture, each cross-attention layer contains two parallel branches: one attending to text embeddings and one attending to image embeddings. Let $A_{\text{text}}^{(l)}$ and $A_{\text{img}}^{(l)}$ denote the attention scores from the text and image branches at layer $l$, respectively. The text/image attention ratio $A$ is computed as:
\begin{equation}
A = \frac{1}{L} \sum_{l=1}^{L} \frac{\|A_{\text{text}}^{(l)}\|_1}{\|A_{\text{img}}^{(l)}\|_1 + \epsilon},
\end{equation}
where $L$ is the number of cross-attention layers in that block and $\epsilon$ is a small constant for numerical stability. A low ratio indicates that the generation is dominated by image conditioning, signaling a potential leakage attack.

\textbf{Detection results.} As shown in Figure~\ref{fig:detection}, both instruction and empty prompts produce abnormally low text/image attention ratios at early diffusion steps, making them easily distinguishable from benign queries. In contrast, our caption-based attack uses natural text queries and yields ratios within the normal range, thereby evading detection. Moreover, while the empty prompt provides a strong shortcut for image leakage, it underperforms our caption-based approach in eliciting reference images, achieving a CLIP image similarity of 0.815 compared to 0.833 for ours (note that we directly compare the similarity between the reference image and the output image, rather than measuring MIA performance). The empty prompt also complicates retrieval segment construction, as the absence of textual context weakens the optimization signal. Together, these results demonstrate that our extraction segment design is both \textit{stealthy} and \textit{effective}.

\section{Additional Experiment Analysis}
\label{app:experiments}
\paragraph{{\rag} improves the utility of both T2I and Q\&A tasks.} As shown in Appendix Tables~\ref{tab:t2i_utility} and~\ref{tab:t2t_utility}, {\rag} substantially improves T2I generation utility (measured by CLIP-I) by over 10\% on the ImageNet100, WikiArt, Dogs, and Cars datasets, and boosts response accuracy on MMQA from 30\% to approximately 55\%. These gains stem from the inherent limitations of parametric knowledge in generative models: (1) fine-grained visual concepts, long-tail entities, and domain-specific details are often poorly learned and underrepresented in the model's parameters; and (2) novel concepts that emerge after the base diffusion model or VLM is trained remain entirely absent from its parametric knowledge. Both cases lead to hallucinated or generic outputs. By retrieving semantically relevant images at inference time, {\rag} grounds generation in concrete visual evidence, allowing the model to faithfully reproduce category-specific details in T2I tasks and to answer visually grounded questions that would otherwise exceed its parametric capacity. This confirms that {\rag} serves as an effective external memory, complementing the model's internal knowledge with up-to-date and fine-grained visual information.

\textbf{Detailed Analysis of Multi-Query Attack Strategies.} We evaluate different variants of attack strategies for multi-query attacks in Appendix Figures~\ref{fig:exp_aggregation} and~\ref{fig:exp_queries}.
Specifically, we observe that different aggregation strategies achieve comparable performance under the standard setting, but the average-score-based approach exhibits much lower performance ($\sim$ 10\% drop in TPR@5\%FPR) under the restrictive setting. We explain that under this more challenging setting, many queries fail to retrieve the test image despite its presence in the database and, therefore, produce noisy scores. The K-means-based strategy mitigates this by aggregating scores from high-confidence clusters, which balances precision with coverage and yields more stable performance across settings. In Figure~\ref{fig:exp_queries}, we show that increasing the number of queries slightly improves attack results, while notably, even the single-query attack achieves desirable performance. 

\section{Additional Details}
\label{app:details1}
\begin{figure*}[t]
    \centering
    \begin{subfigure}[b]{0.32\textwidth}
    \centering
    \includegraphics[width=\textwidth]{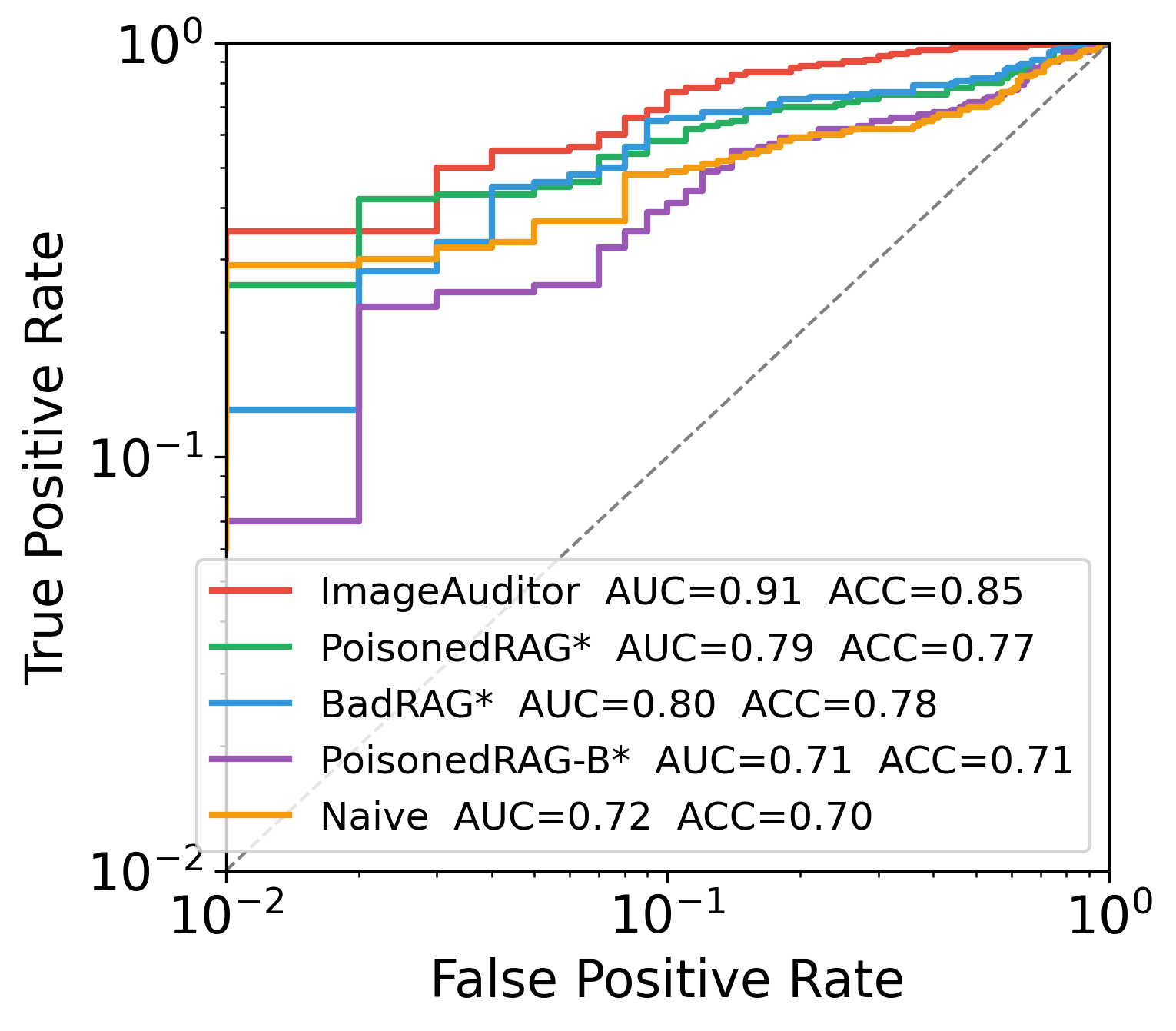}
    \caption{T2I-MSCOCO}
  \end{subfigure}
  \hfill
  \begin{subfigure}[b]{0.32\textwidth}
    \centering
    \includegraphics[width=\textwidth]{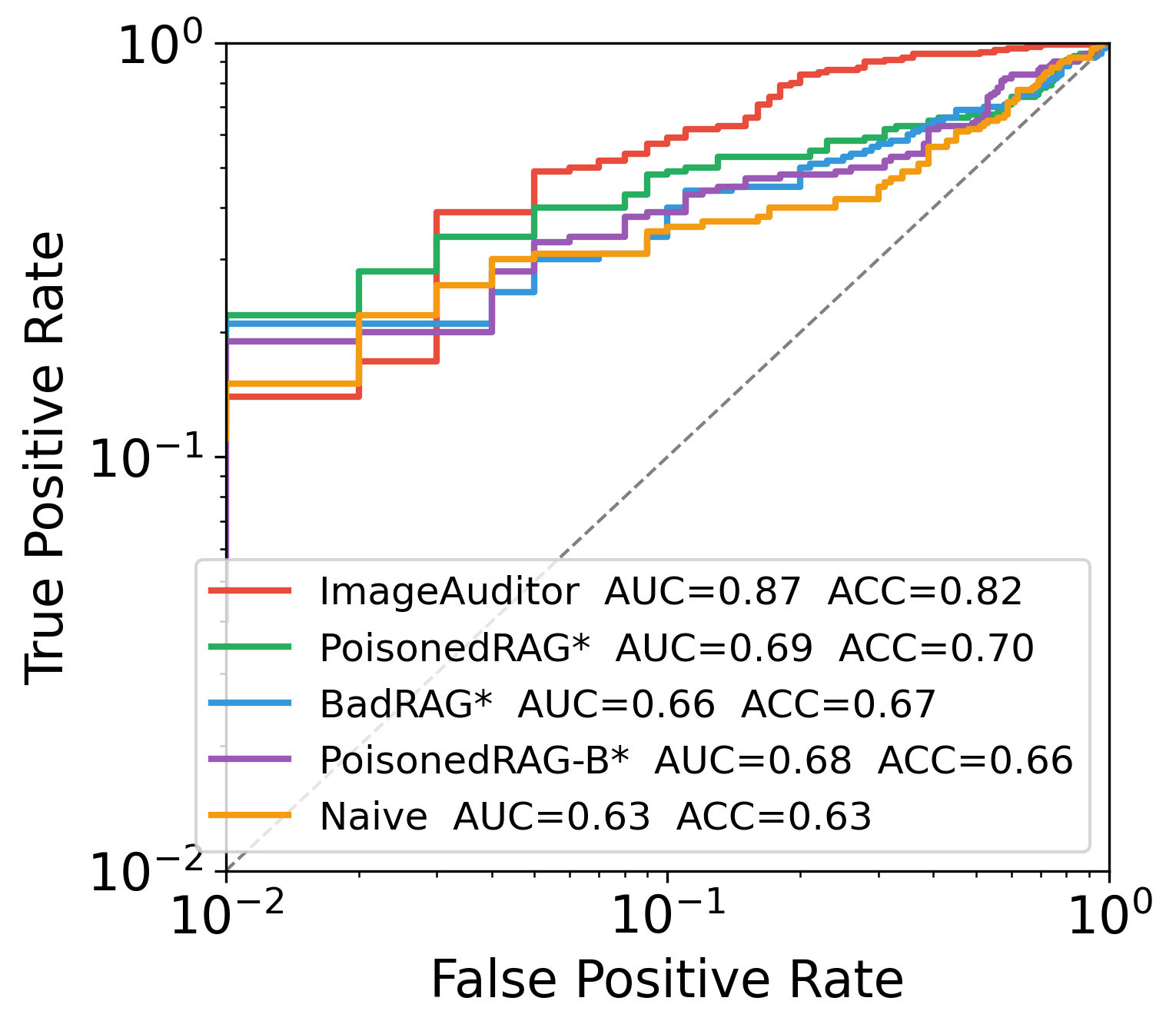}
    \caption{T2I-WikiArt}
  \end{subfigure}
  \hfill
  \begin{subfigure}[b]{0.32\textwidth}
    \centering
    \includegraphics[width=\textwidth]{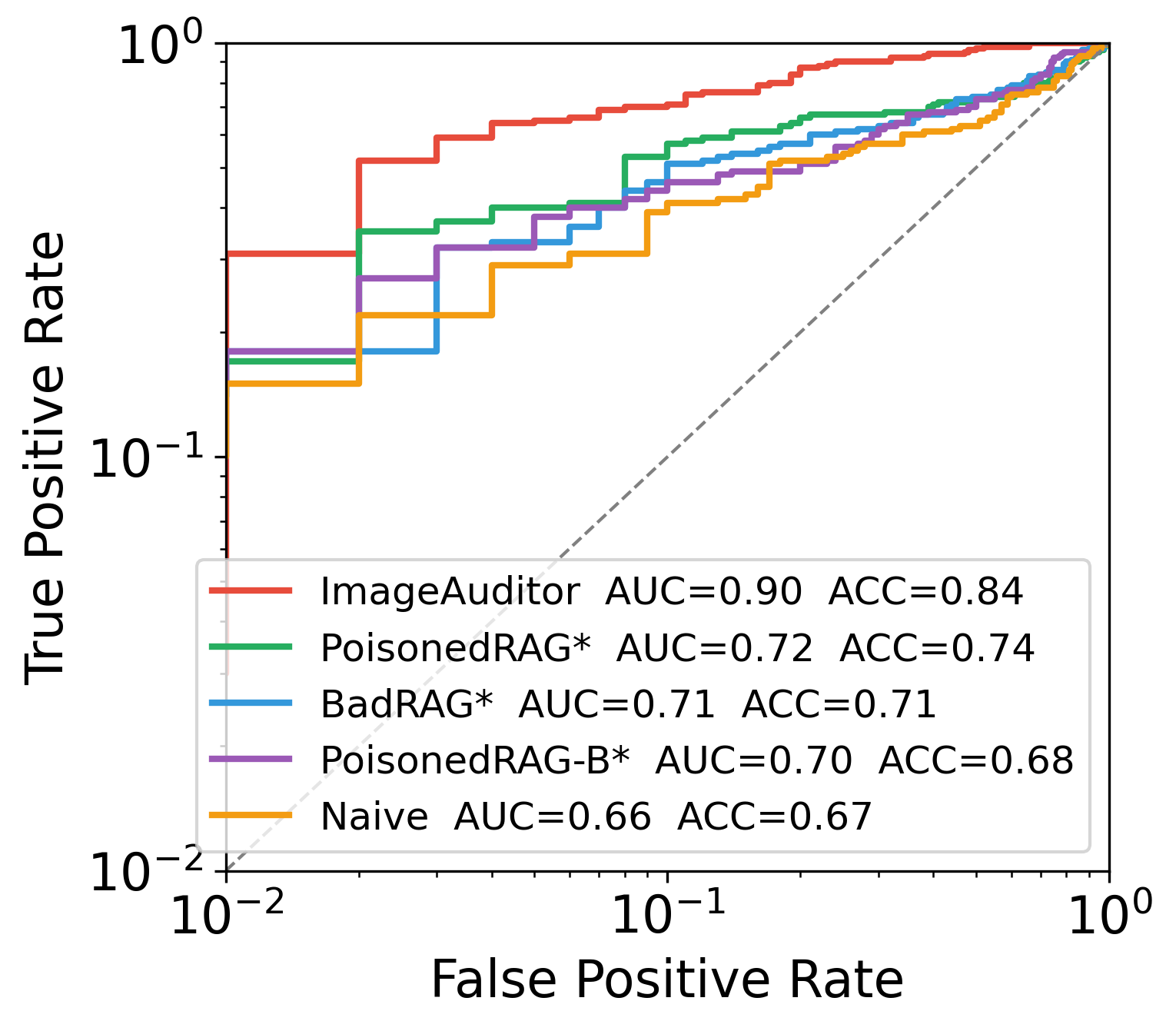}
    \caption{T2I-Stanford Dogs}
    \end{subfigure}
    \hfill
    \caption{TPR–FPR curves of {\name} and adapted baselines on IRAG for T2I generation tasks.}
    \label{fig:exp_t2i_curves}
\end{figure*}

\begin{figure*}[t]
    \centering
    \makebox[\textwidth][c]{%
        \begin{subfigure}[b]{0.32\textwidth}
            \centering
            \includegraphics[width=\textwidth]{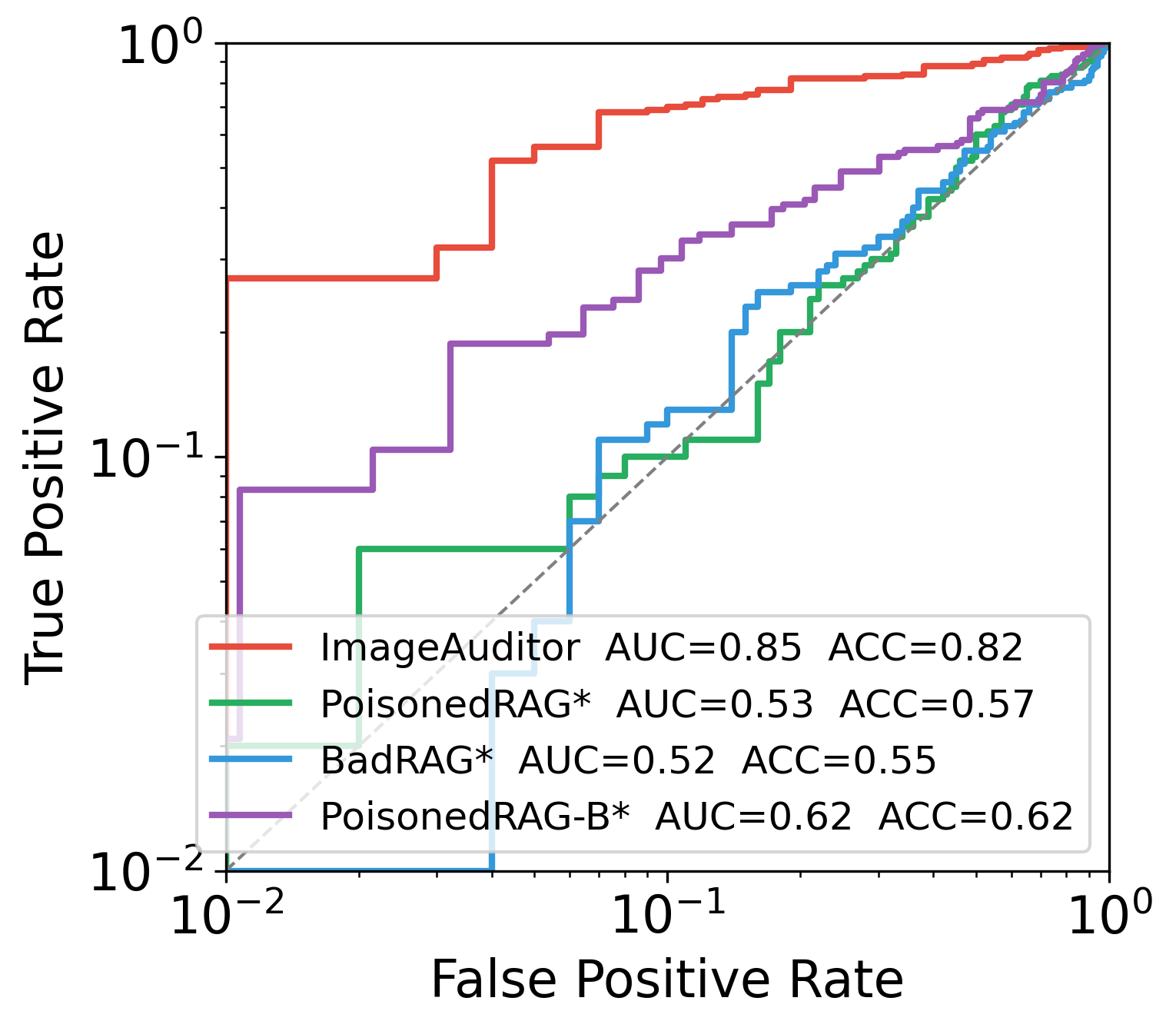}
            \caption{Q\&A-MMQA}
        \end{subfigure}
        \hspace{0.03\textwidth}
        \begin{subfigure}[b]{0.32\textwidth}
            \centering
            \includegraphics[width=\textwidth]{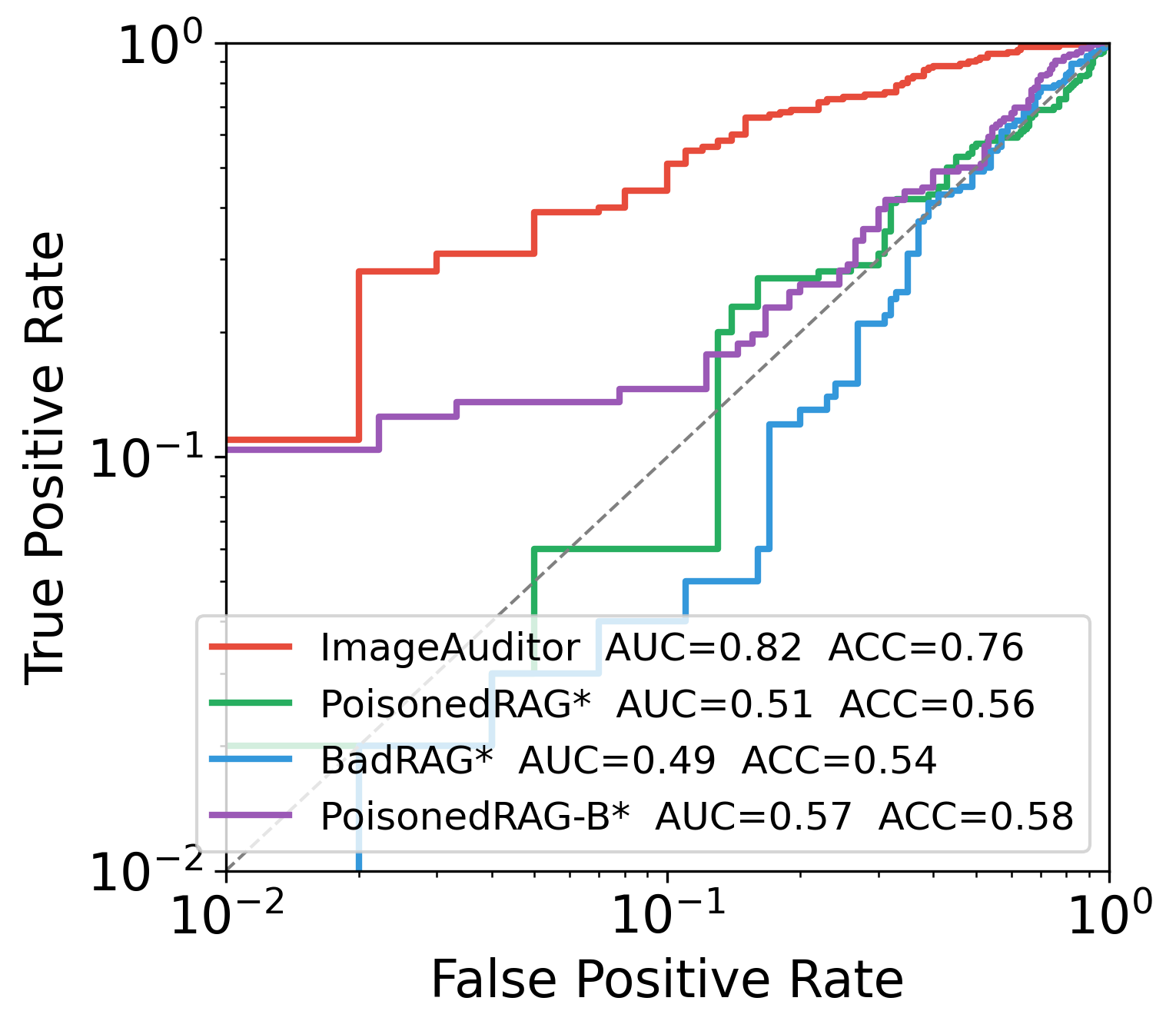}
            \caption{Q\&A-MSCOCO}
        \end{subfigure}
    }
    \caption{TPR--FPR curves of {\name} and adapted baselines on IRAG for Q\&A tasks. Adapted PoisonedRAG and BadRAG fail because the extraction segment—``Describe each retrieved image. Format as 'Image 1:', 'Image 2:', etc.''——provides no semantic alignment with the target image, causing prior attacks to get trapped in local minima and fail to retrieve that image.}
    \label{fig:exp_t2t_curves}
\end{figure*}

\subsection{Description of the Datasets.}
\label{app:dataset}
For both T2I and Q\&A tasks, we construct large-scale image databases by aggregating data from multiple sources, following prior work~\citep{lyu2025realrag,chen2024mllm}. For datasets without official splits, we partition the data into member and non-member sets, using the member split to build the image database. 

\textbf{T2I datasets.}
\textit{(1) MSCOCO (2017)}~\citep{lin2014microsoft} is a large-scale image-caption dataset of real-world photographic images, containing 118K training images and 5K validation images.
\textit{(2) ImageNet-100} is a 100-class subset of ImageNet~\citep{russakovsky2015imagenet} comprising $\sim$130K images across diverse object categories. For controlled experiments, we use a 50-class subset by default and vary the number of classes (from 1 to 100) in ablation studies.
\textit{(3) WikiArt}~\citep{phillips2011wiki} contains 81,444 paintings spanning 27 styles and 45 genres from over 2,000 artists across centuries of art history. We exclude paintings labeled as unknown artists, yielding a filtered dataset of $\sim$39K images. We then split the dataset by artist at a 4:1 ratio.
\textit{(4) Stanford Cars}~\citep{krause20133d} consists of 16,185 images of 196 car models, with an official 8,144/8,041 train/test split.
\textit{(5) Stanford Dogs}~\citep{dataset2011novel} contains 20,580 images of 120 dog breeds, with an official 12,000/8,580 train/test split.
\textit{(6) CelebA-HQ}~\citep{karras2017progressive} is a high-quality facial dataset containing 30,000 face images at 1024$\times$1024 resolution. We split the dataset at a 4:1 ratio. In total, we construct an image database of 259{,}057 images spanning all six datasets. Together, they provide concepts covering \textit{real-world scenes}, \textit{fine-grained and rare objects}, \textit{artistic styles}, and \textit{high-quality faces}, etc. 

\textbf{Q\&A dataset.} \textit{(1) MMQA} (MultiModalQA)~\citep{talmor2021multimodalqa} is a multi-modal question answering benchmark requiring reasoning across text, tables, and images. We use the complete image partition of the dataset, containing 58,075 images. Due to the lack of official splits, we randomly sample 1,000 images as the non-member set. \textit{(2) MSCOCO} is also used in RagVL~\citep{chen2024mllm} as a Q\&A dataset to evaluate text-to-image retrieval performance over a large-scale image database.

\textbf{Utility evaluation.} The T2I database covers diverse visual concepts spanning real scenes, fine-grained and rare objects, and artistic styles. For datasets with clear class labels, we randomly sample 20 classes and generate 10 prompts per class—for example, \textit{A photo of a \{class\} sitting on the ground''} for Stanford Dogs, and \textit{A modern artwork influenced by \{class\}''} for WikiArt. This allows us to evaluate whether generated images faithfully capture the visual concepts from each class and to compare generators with and without {\rag}. For MSCOCO and CelebA-HQ, we randomly sample 200 prompts following their respective distributions and compare the generated image distributions using FID and CLIP scores. For MMQA, which supplies visual knowledge for knowledge-dependent question answering, we follow the same setup as AQUA and measure response accuracy to compare generators with and without {\rag}.

\begin{table*}[t]
\centering
\caption{Main results of {\name} on {\rag} for Q\&A tasks. Following the literature~\cite{chen2024mllm}, the system uses CLIP ViT-L/14-336 for retrieval and Qwen2.5-VL-7B-Instruct for generation. All methods share the same extraction segment (``Describe each retrieved image...'') and differ only in their retrieval segment, so the results reflect optimizer quality. PoisonedRAG-B$^*$ outperforms the two gradient-based variants as it avoids noisy gradient estimation in this challenging setting.}
\label{tab:t2t_attack_results}
\small
\setlength{\tabcolsep}{4.5pt} 
\renewcommand{\arraystretch}{0.9}

\begin{tabular}{l ccc ccc}
\toprule
\textbf{Attack Method}
& \multicolumn{3}{c}{\textbf{MMQA}}
& \multicolumn{3}{c}{\textbf{MSCOCO}} \\
\cmidrule(lr){2-4} \cmidrule(lr){5-7}

& AUC & ACC & {\tprfpr}
& AUC & ACC & {\tprfpr} \\
\midrule

Na\"ive
& \multicolumn{3}{c}{N/A}
& \multicolumn{3}{c}{N/A} \\
\cmidrule(lr){1-7}

PoisonedRAG-B$^*$
& 0.62 & 0.62 & 0.19
& 0.57 & 0.59 & 0.14 \\

PoisonedRAG$^*$
& 0.53 & 0.57 & 0.06
& 0.51 & 0.56 & 0.06 \\

BadRAG$^*$
& 0.52 & 0.55 & 0.04
& 0.49 & 0.54 & 0.03 \\

{\name}
& \textbf{0.85} & \textbf{0.82} & \textbf{0.56}
& \textbf{0.82} & \textbf{0.76} & \textbf{0.39} \\
\bottomrule
\end{tabular}
\end{table*}

\begin{table*}[t]
\centering
\begin{minipage}{0.49\textwidth}
\centering
\caption{Stability of our evaluation. MSCOCO and MMQA are used for T2I and Q\&A tasks.}
\label{tab:stability}
\small
\setlength{\tabcolsep}{3.3pt}
\renewcommand{\arraystretch}{0.9}
\begin{tabular}{lc|ccc}
\toprule
\textbf{Task} & \textbf{\# Tests} & \textbf{AUC} & \textbf{ACC} & \textbf{\tprfpr} \\
\midrule
\multirow{2}{*}{T2I task}
    & Default & 0.91 & 0.85 & 0.55 \\
    & 800 & 0.91 & 0.85 & 0.56 \\
\midrule
\multirow{2}{*}{Q\&A task}
    & Default & 0.85 & 0.82 & 0.56 \\
    & 800 & 0.85 & 0.80 & 0.49 \\
\bottomrule
\end{tabular}
\end{minipage}
\hfill
\begin{minipage}{0.49\textwidth}
\centering
\caption{Attack evaluation on T2I tasks under both per-dataset and combined settings.}
\label{tab:full_mia}
\small
\setlength{\tabcolsep}{3.3pt}
\renewcommand{\arraystretch}{0.9}
\begin{tabular}{cc|ccc}
\toprule
\textbf{Setting} & \textbf{Dataset} & \textbf{AUC} & \textbf{ACC} & \textbf{\tprfpr} \\
\midrule
\multirow{3}{*}{Separate} & MSCOCO    & 0.91 & 0.85 & 0.55 \\
& WikiArt & 0.87 & 0.82 & 0.49 \\
& Dogs    & 0.90 & 0.84 & 0.65 \\
\midrule
Combined & Full    & 0.87 & 0.79 & 0.46 \\
\bottomrule
\end{tabular}
\end{minipage}
\end{table*}

\subsection{Description of the Generation Settings}
\label{app:details2}
For T2I tasks, we use four generators that span different architectures and conditioning mechanisms:
\textit{(1) SDXL + IP-Adapter}~\citep{podell2023sdxl,ye2023ip}: We pair SDXL with the ViT-H IP-Adapter Plus variant (\texttt{ip-adapter-plus\_sdxl\_vit-h}) and use an IP-Adapter scale of 0.6, which provides a balanced setting between image and text conditioning. Images are generated at 1024$\times$1024 resolution with 30 denoising steps, following the default setting of SDXL. Empirically, ImageAuditor maintains AUROC above 80\% across practical scale values ($\geq 0.5$); lower scales weaken retrieval influence and are rarely used in IRAG systems, as they reduce the benefit of retrieval augmentation. \textit{(2) SD1.5 + IP-Adapter}~\citep{rombach2022high,ye2023ip}: We pair Stable Diffusion 1.5 with its corresponding IP-Adapter Plus variant and use an IP-Adapter scale of 0.6. Images are generated at 512$\times$512 resolution with 30 denoising steps, following the default setting of SD1.5.
\textit{(3) Kandinsky v2.2}~\citep{razzhigaev2023kandinsky}: We use Kandinsky v2.2, a latent diffusion model with a CLIP-based image prior that natively supports image-conditioned generation. Images are generated at 512$\times$512 resolution with 100 denoising steps, using an image weight of 0.7, a guidance scale of 4.0, a prior guidance scale of 4.0, and 25 prior inference steps.
\textit{(4) SDXL + Conceptrol}~\citep{he2025conceptrol}: We use SDXL augmented with Conceptrol, an advanced image conditioning method. Images are generated at 1024$\times$1024 resolution with 30 denoising steps, using a condition scale of 1.0, a guidance scale of 6.0, and a warmup ratio of 0.2.

For Q\&A tasks, we use two vision-language models (VLMs):
\textit{(1) Qwen2.5-VL}~\citep{wang2024qwen2}: We experiment with Qwen2.5-VL, a vision-language model from the Qwen family with native dynamic resolution support. \textit{(2) LLaVA-1.6-Mistral-7B}~\citep{liu2023visual}: We also use LLaVA-1.6-Mistral-7B, a 7B-parameter vision-language model built on Mistral that supports dynamic high-resolution image inputs. All other hyperparameters follow the default settings from each model's official implementation.

\begin{table*}[t]
\centering
\caption{{\rag} improves the utility of text-to-image generation. Following~\cite{lyu2025realrag}, we use CLIP-I as a metric of concept-level semantic alignment, which measures the similarity between the generated image and the corresponding class prototype embedding.}
\label{tab:t2i_utility}
\small
\setlength{\tabcolsep}{3.2pt}
\renewcommand{\arraystretch}{0.9}
\begin{tabular}{llccccccccc}
\toprule
\textbf{Generator} & \textbf{RAG}
& \multicolumn{3}{c}{\textbf{WikiArt}}
& \multicolumn{3}{c}{\textbf{ImageNet100}}
& \multicolumn{3}{c}{\textbf{Stanford Dogs}} \\
\cmidrule(lr){3-5}
\cmidrule(lr){6-8}
\cmidrule(lr){9-11}
& 
& CLIP-I & CLIP-T & FID
& CLIP-I & CLIP-T & FID
& CLIP-I & CLIP-T & FID \\
\midrule
\multirow{2}{*}{SDXL+IP-Adapter}
    & $\times$      & 0.453 & \textbf{0.314} & 330 & 0.649 & 0.343 & 145 & 0.430 & \textbf{0.324} & 235 \\
    & $\checkmark$  & \textbf{0.522} & 0.296 & \textbf{292} & \textbf{0.744} & \textbf{0.358} & \textbf{103} & \textbf{0.623} & 0.269 & \textbf{138} \\
\midrule
\multirow{2}{*}{SD1.5+IP-Adapter}
    & $\times$      & 0.419 & 0.292 & 340 & 0.617 & 0.312 & 189 & 0.412 & \textbf{0.286} & 243 \\
    & $\checkmark$  & \textbf{0.539} & \textbf{0.295} & \textbf{297} & \textbf{0.732} & \textbf{0.321} & \textbf{108} & \textbf{0.621} & 0.264 & \textbf{135} \\
\midrule
\midrule
\textbf{Generator} & \textbf{RAG}
& \multicolumn{3}{c}{\textbf{Cars}}
& \multicolumn{3}{c}{\textbf{MSCOCO}}
& \multicolumn{3}{c}{\textbf{CelebA-HQ}} \\
\cmidrule(lr){3-5}
\cmidrule(lr){6-8}
\cmidrule(lr){9-11}
& 
& CLIP-I & CLIP-T & FID
& CLIP-I & CLIP-T & FID
& CLIP-I & CLIP-T & FID \\
\midrule
\multirow{2}{*}{SDXL+IP-Adapter}
    & $\times$      & 0.618 & 0.333 & 160 & \slash & 0.362 & 190 & \slash & 0.365 & 171 \\
    & $\checkmark$  & \textbf{0.688} & \textbf{0.334} & \textbf{112} & \slash & \textbf{0.364} & \textbf{178} & \slash & \textbf{0.375} & \textbf{117} \\
\midrule
\multirow{2}{*}{SD1.5+IP-Adapter}
    & $\times$      & 0.608 & 0.311 & 124 & \slash & 0.343 & 180 & \slash & 0.351 & 112 \\
    & $\checkmark$  & \textbf{0.684} & \textbf{0.319} & \textbf{111} & \slash & \textbf{0.347} & \textbf{175} & \slash & \textbf{0.364} & \textbf{96} \\
\bottomrule
\end{tabular}
\end{table*}

\begin{table*}[t]
\centering
\caption{{\rag} improves the utility of question answering. We report accuracy (ACC) on MMQA across different numbers of retrieved passages.}
\label{tab:t2t_utility}
\small
\setlength{\tabcolsep}{3.3pt}
\renewcommand{\arraystretch}{0.9}
\begin{tabular}{lccc}
\toprule
\textbf{Generator} & None & \textbf{$K=2$} & \textbf{$K=3$} \\
\midrule
Qwen2.5-7B-Instruct & 0.314 & \textbf{0.529} & 0.495 \\
LLaVA-v1.6-Mistral-7B   & 0.359 & \textbf{0.575} & 0.573 \\
\bottomrule
\end{tabular}
\end{table*}

\begin{figure*}[t]
    \centering
    \includegraphics[width=0.6\textwidth]{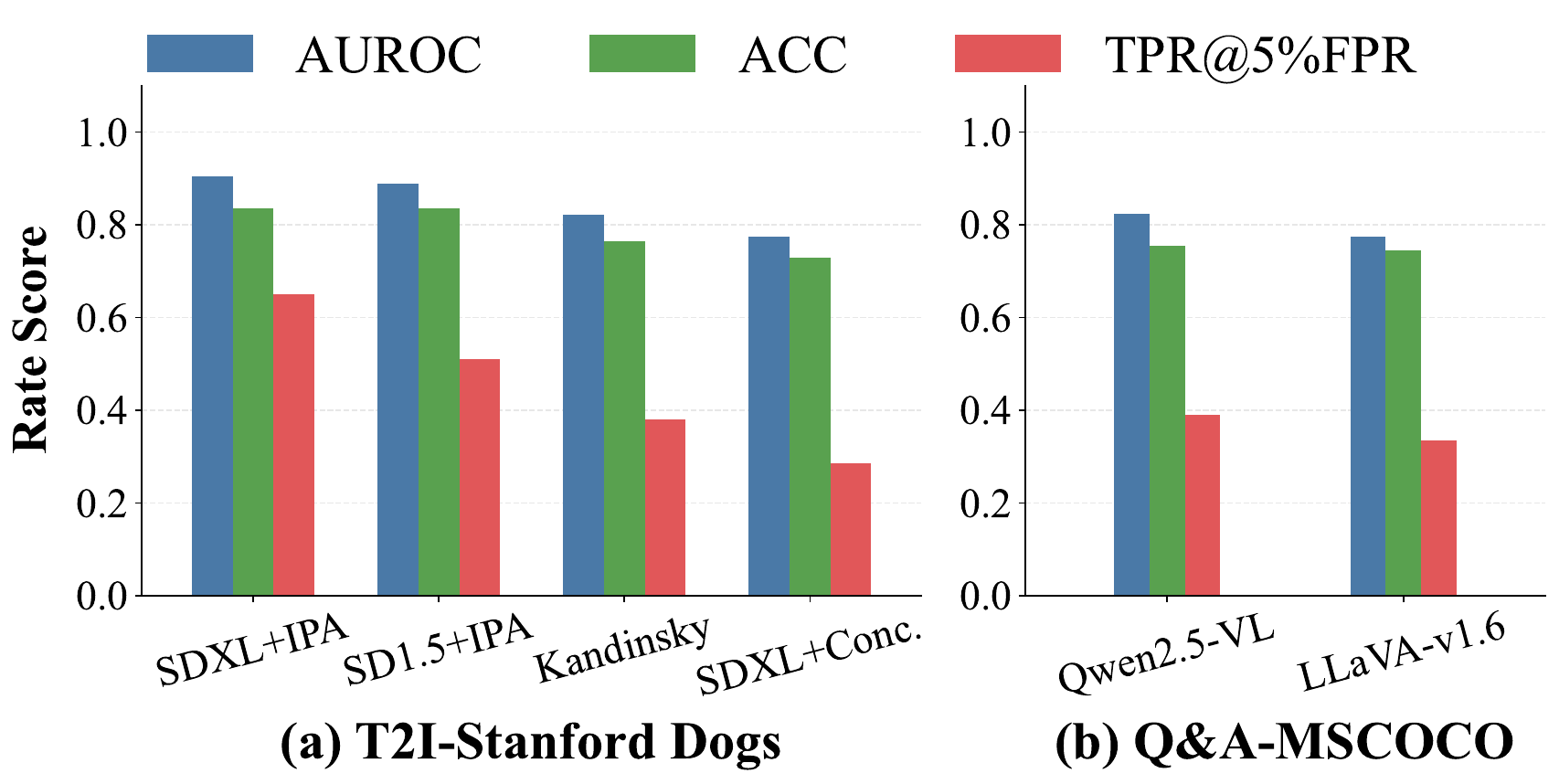}
    \caption{{\name} remains effective across different generators for both T2I and Q\&A tasks. SDXL+Conceptrol reduces TPR@5\%FPR due to its defensive and more creative generation behavior. However, it still achieves around 30\% of TPR@5\%FPR, which significantly outperforms random guessing (5\%) and is considered an effective attack under standard MIA literature.}
    \label{fig:exp_generator2}
\end{figure*}

\begin{table*}[t]
\centering
\caption{Impact of extraction segment type on {\name} against IRAG for the T2I task. BLIP and Qwen-VL captions yield stronger attacks, as better alignment between the image and the extraction segment more effectively elicits copy-like behavior from T2I generators.}
\label{tab:exp_extraction}
\small
\setlength{\tabcolsep}{3pt}
\renewcommand{\arraystretch}{0.9}
\begin{tabular}{lccc ccc ccc}
\toprule
\multirow{2}{*}{\shortstack{\textbf{Extraction Segment}\\ \textbf{Type}}}
& \multicolumn{3}{c}{\textbf{MSCOCO}}
& \multicolumn{3}{c}{\textbf{WikiArt}} 
& \multicolumn{3}{c}{\textbf{Stanford Dogs}} \\
\cmidrule(lr){2-4} 
\cmidrule(lr){5-7} 
\cmidrule(lr){8-10}
& AUC & ACC & {\tprfpr}
& AUC & ACC & {\tprfpr}
& AUC & ACC & {\tprfpr} \\
\midrule
Generic Cap. & 0.71 & 0.70 & 0.38 & 0.75 & 0.72 & 0.39 & 0.83 & 0.76 & 0.49 \\
BLIP Cap.    & 0.86 & 0.82 & 0.42 & \textbf{0.87} & 0.80 & \textbf{0.54} & 0.88 & 0.83 & 0.51 \\
Qwen-VL Cap.  & \textbf{0.91} & \textbf{0.85} & \textbf{0.55} & \textbf{0.87} & \textbf{0.82} & 0.49 & \textbf{0.90} & \textbf{0.84} & \textbf{0.65} \\
\bottomrule
\end{tabular}
\end{table*}

\begin{table}[t]
\centering
\caption{The shadow dataset improves results when the target embedding space is unknown. Here, ViT-L/14 is used as the shadow embedding model.}
\label{tab:exp_shadow}
\small
\setlength{\tabcolsep}{3pt}
\renewcommand{\arraystretch}{0.9}
\begin{tabular}{c|ccccccccc}
\toprule
\textbf{Type} & \multicolumn{3}{c}{\textbf{MSCOCO}} & \multicolumn{3}{c}{\textbf{WikiArt}}
& \multicolumn{3}{c}{\textbf{Stanford Dogs}} \\
\cmidrule(lr){2-4} \cmidrule(lr){5-7}   \cmidrule(lr){8-10}
& \textbf{AUC} & \textbf{ACC} & \textbf{\tprfpr} & \textbf{AUC} & \textbf{ACC} & \textbf{\tprfpr} & \textbf{AUC} & \textbf{ACC} & \textbf{\tprfpr} \\
\midrule
No &0.75 &0.70 &0.31 & 0.64 & 0.64 & 0.18 & 0.72 & 0.68 & 0.26 \\
Yes & \textbf{0.78} & \textbf{0.75} & \textbf{0.32} & \textbf{0.70} & \textbf{0.67} & \textbf{0.34} & \textbf{0.75} & \textbf{0.70} & \textbf{0.33} \\
\bottomrule
\end{tabular}
\end{table}

\begin{figure*}[t]
    \centering
    \begin{subfigure}[t]{0.49\textwidth}
        \centering
        \includegraphics[width=\linewidth]{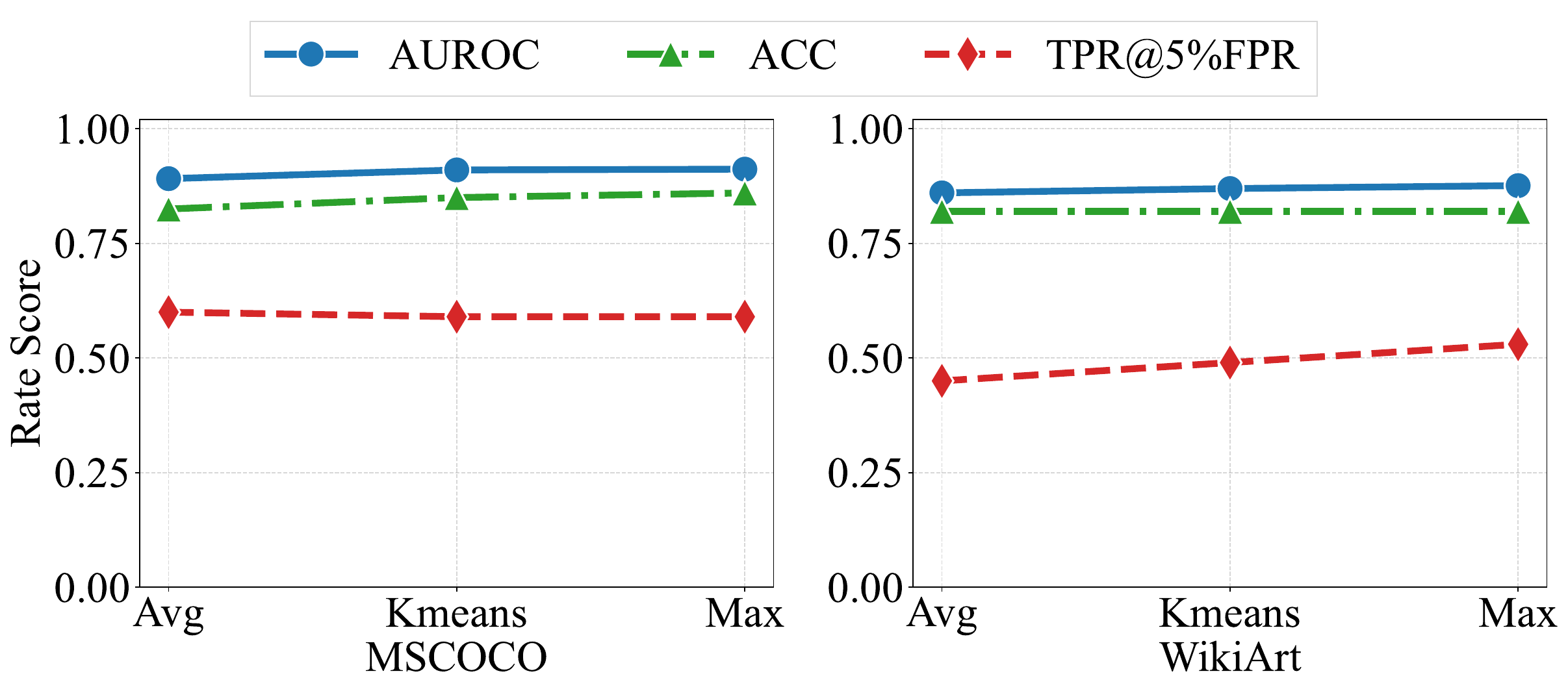}
        \caption{with access to ViT-H/14 (standard setting)}
    \end{subfigure}
    \hfill
    \begin{subfigure}[t]{0.49\textwidth}
        \centering
        \includegraphics[width=\linewidth]{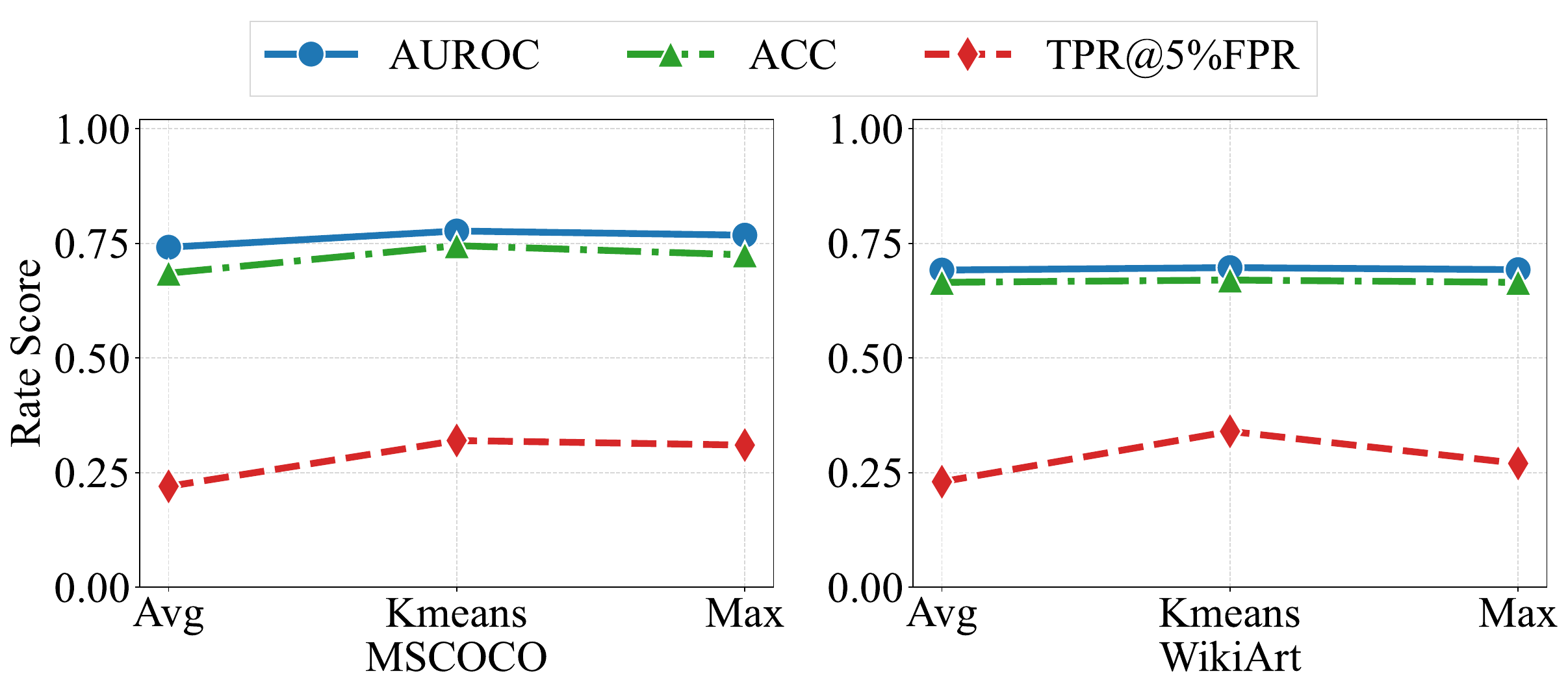}
        \caption{with access to ViT-L/14 (restrictive setting)}
    \end{subfigure}
    \hfill
\caption{K-means-based aggregation yields the most stable attack results across various settings.}
\label{fig:exp_aggregation}
\end{figure*}

\begin{figure*}[t]
    \centering
    \begin{minipage}[t]{0.49\textwidth}
        \centering
        \includegraphics[width=\linewidth]{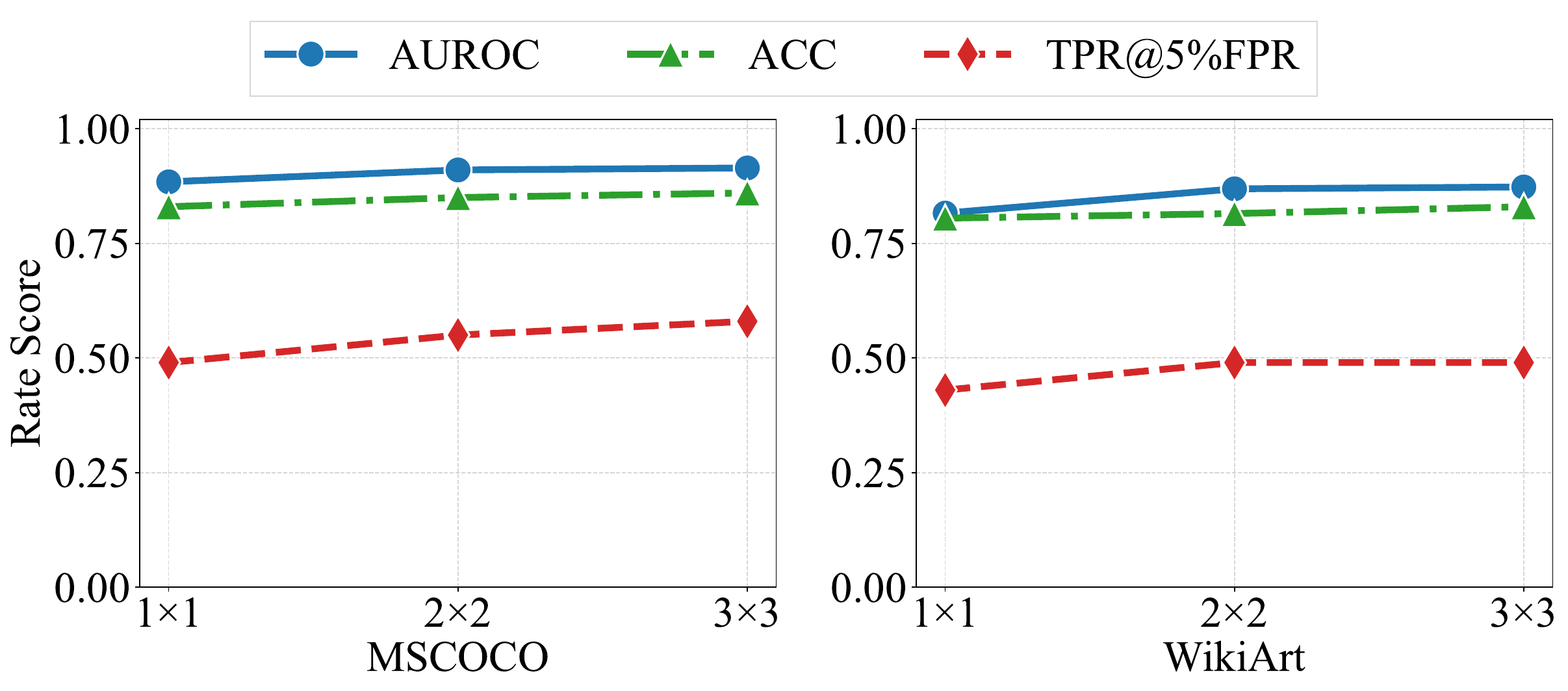}
        \caption{Increasing the number of queries ($M \times N$) per test case slightly improves the attack performance of {\name}. Our attack remains effective with a single query due to the RGPO mechanism.}
        \label{fig:exp_queries}
    \end{minipage}
    \hfill
    \begin{minipage}[t]{0.49\textwidth}
        \centering
        \includegraphics[width=\linewidth]{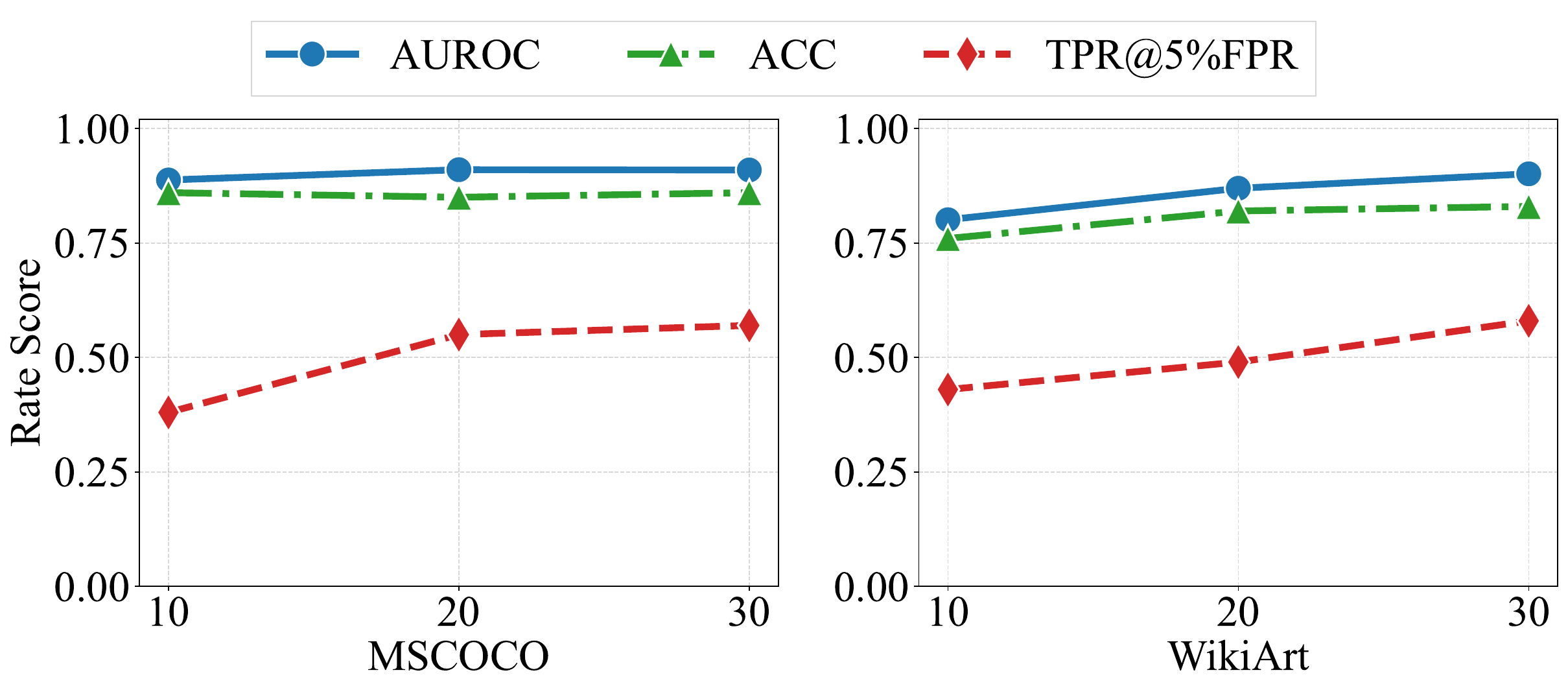}
        \caption{Increasing the number of RGPO iterations slightly improves the results (efficiency vs. effectiveness trade-off), and {\name} remains effective even with a small number of iterations (e.g., 10).}
        \label{fig:exp_iterations}
    \end{minipage}
\end{figure*}

\begin{table*}[t]
\centering
\caption{In both standard and restrictive settings, a stronger scoring model improves attack performance, as it more accurately captures fine-grained image semantics.}
\label{tab:scoring_model}
\small
\setlength{\tabcolsep}{3.3pt}
\renewcommand{\arraystretch}{0.9}
\begin{tabular}{llcccccc}
\toprule
\textbf{Dataset}& \textbf{Scoring model}
& \multicolumn{3}{c}{\textbf{with access to ViT-H/14}}
& \multicolumn{3}{c}{\textbf{with access to ViT-L/14}} \\
\cmidrule(lr){3-5} \cmidrule(lr){6-8}
& & AUC & ACC & {\tprfpr}
& AUC & ACC & {\tprfpr} \\
\midrule
\multirow{2}{*}{MSCOCO} & EvaCLIP & 0.87 & 0.85 & 0.32 & 0.78 & 0.75 & 0.32 \\
& CLIP ViT-H/14 & \textbf{0.91} & \textbf{0.85} & \textbf{0.55} & \textbf{0.82} & \textbf{0.78} & \textbf{0.39} \\
\midrule 
\multirow{2}{*}{WikiArt} & EvaCLIP & 0.81 & 0.77 & 0.32 &0.68 & 0.68 & 0.27 \\
& CLIP ViT-H/14 & \textbf{0.87} & \textbf{0.82} & \textbf{0.49} & \textbf{0.69} & \textbf{0.71} & \textbf{0.41} \\
\midrule
\multirow{2}{*}{Dogs} & EvaCLIP & 0.88 & 0.82 & 0.53 & 0.75 & 0.70 & 0.33 \\
& CLIP ViT-H/14 & \textbf{0.90} & \textbf{0.84} & \textbf{0.65} & \textbf{0.77} & \textbf{0.73} & \textbf{0.41} \\
\bottomrule
\end{tabular}
\end{table*}

\subsection{Description of the Evaluation Metrics}
\label{app:metrics}
For attack evaluation, we use standard MIA metrics, including AUROC, ACC, and TPR@5\%FPR. AUROC measures the area under the ROC curve and captures the overall separability between members and non-members across all thresholds. ACC reports the classification accuracy at the optimal decision threshold. TPR@5\%FPR measures the true positive rate when the false positive rate is constrained to 5\%, reflecting performance under a strict false-alarm budget. We additionally report the Retrieval Success Rate (RSR) to measure the attack's effectiveness at retrieving member images. For utility evaluation, we report CLIP-T, CLIP-I (key metric for class-labeled dataset), and FID for T2I tasks and response accuracy for Q\&A tasks, following the literature~\cite{lyu2025realrag,chen2024mllm}.

\subsection{Description of the Attack Baselines}
\label{app:baselines}
Traditional text-RAG MIAs~\cite{naseh2025riddle,liu2025mask} are not directly applicable to our setting due to the incompatibility of T2I generators with question-answering tasks and the inability to embed images into text queries. We therefore adapt these baselines following the query decomposition paradigm of \textit{\name}. The naïve baseline includes only the extraction segment: we generate candidate captions and select those with high reward (e.g., similarity to the test image), following our framework. Since these captions semantically describe the test image, this baseline can still retrieve some member images in the T2I setting, providing a meaningful comparison point.

We additionally adapt PoisonedRAG-B~\citep{zou2025poisonedrag}, PoisonedRAG~\citep{zou2025poisonedrag}, and BadRAG~\citep{xue2024badrag} as baseline attacks. All three are originally designed for text-RAG poisoning, and we adapt them by treating retrieval segment construction as analogous to their adversarial passage optimization. Specifically, we apply their respective objectives to construct retrieval segments for each test image, while keeping the extraction segment and scoring rule identical to {\name} for a fair comparison. The three methods differ primarily in their optimization objective.

\textit{(1) PoisonedRAG-B}~\citep{zou2025poisonedrag} is the gradient-free variant of PoisonedRAG. Rather than optimizing $T_{\text{ret}}$ via token-level gradients, it directly constructs the retrieval segment by extracting content from the target sample. In our case, we use the generated image caption as $T_{\text{ret}}$. In the T2I setting, this differs from the Na\"ive baseline in that Na\"ive uses no retrieval segment, whereas PoisonedRAG-B$^{*}$ employs two separate captions---one as the extraction segment and another as the retrieval segment.

\textit{(2) PoisonedRAG}~\citep{zou2025poisonedrag} employs HotFlip~\citep{ebrahimi2018hotflip} to optimize the retrieval segment $T_{\text{ret}}$ such that the embedding of the full query is close to that of the target image. Specifically, given a target embedding $\mathbf{e}_t$, a query encoder $E(\cdot)$, and a fixed extraction segment $T_{\text{ext}}$, the optimization objective is:
\begin{equation}
    T_{\text{ret}}^* = \arg\max_{T_{\text{ret}}} \; \text{sim}\bigl(E(T_{\text{ext}} \oplus T_{\text{ret}}),\, \mathbf{e}_t\bigr),
\end{equation}
where $\text{sim}(\cdot,\cdot)$ denotes cosine similarity and $\oplus$ denotes concatenation. At each iteration, HotFlip selects a random token position and replaces the token with the candidate yielding the largest gradient-based score improvement.

\textit{(3) BadRAG}~\citep{xue2024badrag} adopts a contrastive optimization objective (COP) that distinguishes the target from a set of shadow (negative) samples. Let $\mathbf{q} = E(T_{\text{ext}} \oplus T_{\text{ret}})$ denote the query embedding and $\{\mathbf{e}_n^{(i)}\}_{i=1}^{N}$ denote shadow embeddings drawn from the shadow set. The InfoNCE-style objective is:
\begin{equation}
    T_{\text{ret}}^* = \arg\max_{T_{\text{ret}}} \; \log \frac{\exp(\text{sim}(\mathbf{q}, \mathbf{e}_t) / \tau)}{\exp(\text{sim}(\mathbf{q}, \mathbf{e}_t) / \tau) + \sum_{i=1}^{N} \exp(\text{sim}(\mathbf{q}, \mathbf{e}_n^{(i)}) / \tau)},
\end{equation}
where $\tau$ is a temperature hyperparameter. This formulation encourages the retrieval segment to be selectively retrieved for the target while avoiding retrieval for negatives. The optimization is solved using HotFlip-style token-level updates.

\subsection{Hyperparameter Sensitivity}
\label{app:hyper}
The attacker is allowed $M \times N$ queries per test image, where $M$ is the number of extraction segments and $N$ is the number of retrieval segments per extraction segment. We set $M = N = 2$ by default to balance attack effectiveness against query and optimization costs. Using multiple extraction and retrieval segments reduces sensitivity to any single choice. As shown in Figure~\ref{fig:exp_queries}, larger values yield only marginal gains and even a single query (i.e., $M=N=1$) already achieves competitive performance. For the extraction segment, we use Qwen-VL to generate and rank 5 candidate captions in the T2I setting, and manually design structured prompts in the Q\&A setting. For the retrieval segment, we set the token length to 8, which is sufficient for the attack to saturate (see Figure~\ref{fig:exp_rag_abl}(c)). We further sample a small shadow set of 32 images (disjoint from the database) for the contrastive reward, which is practical to collect from public sources. We optimize each retrieval segment for 20 iterations in the T2I setting and 80 iterations in the Q\&A setting, due to differences in extraction prompt designs. As shown in Figure~\ref{fig:exp_iterations}, our designs provide a favorable efficiency--effectiveness trade-off, with diminishing returns beyond this point. The remaining RGPO parameters take standard values from the policy-optimization literature: $\tau = 1$ (a neutral temperature), $\rho = 0.2$ (a common choice in importance sampling methods), and $\beta = 0.5$ (a balanced adaptation rate). We found these defaults to be robust across all datasets, and therefore did not perform dedicated sweeps over them.

\subsection{Description of the Attack Prompts.} 

We use the following prompts in our attack pipeline: 

\begin{tcolorbox}[title=Prompt for Extraction Segment Generation in the T2I Setting, colback=gray!5, colframe=gray!80!black, boxsep=2pt, left=2pt, right=2pt, top=2pt, bottom=2pt, before skip=5pt, after skip=5pt, fonttitle=\normalfont\bfseries, fontupper=\small, breakable]
You are an image captioning system.
\vspace{0.5em}
Given an input image, generate 5 concise and diverse captions.
\vspace{0.5em}
\textbf{Requirements:}
\begin{itemize}[leftmargin=*, noitemsep, topsep=0pt]
\item Write all captions in English only. Do not use any other language.
\item Each caption should describe the main scene and key objects.
\item Keep each caption short and natural (similar to MSCOCO style).
\item Use different wording and focus on slightly different aspects (e.g., objects, actions, relationships, or context).
\item Avoid repeating the same sentence structure or phrases.
\item Do not include unnecessary details or long descriptions.
\end{itemize}
\vspace{0.5em}
\textbf{Output format:}
\begin{verbatim}
1. ...
2. ...
3. ...
4. ...
5. ...
\end{verbatim}
\end{tcolorbox}

\begin{tcolorbox}[title=Extraction Segment in the Q\&A Setting, colback=gray!5, colframe=gray!80!black, boxsep=2pt, left=2pt, right=2pt, top=2pt, bottom=2pt, before skip=5pt, after skip=5pt, fonttitle=\normalfont\bfseries, fontupper=\small, breakable]
Describe each retrieved image. Format as `Image 1:', `Image 2:', etc.
\end{tcolorbox}

\begin{tcolorbox}[title=Prompt for Details Extraction in the Q\&A Setting, colback=gray!5, colframe=gray!80!black, boxsep=2pt, left=2pt, right=2pt, top=2pt, bottom=2pt, before skip=5pt, after skip=5pt, fonttitle=\normalfont\bfseries, fontupper=\small, breakable]
You are an image captioning system.
\vspace{0.5em}

Given an input image, write one detailed natural caption.
\vspace{0.5em}

\textbf{Requirements:}
\begin{itemize}[leftmargin=*, noitemsep, topsep=0pt]
\item Describe visible objects, text, scene, attributes, and layout when helpful.
\item Stay faithful to the image.
\item Do not speculate beyond what is visible.
\item Do not output numbering, explanations, or extra text.
\item Output exactly one caption.
\end{itemize}
\end{tcolorbox}


\section{Discussion}
\label{app:discussion}
\subsection{Generalization to multi-modal inputs.} 
Our attack also extends to {\rag} systems that take image-text pairs as input. Specifically, we can use the image as the retrieval segment and the entire text as the extraction segment, which will further improves attack performance due to the continuous nature of the pixel space. More broadly, this decomposition applies to other multimodal RAG variants, suggesting that the vulnerability is fundamental to cross-modal retrieval.

\subsection{Responsible Deployment of {\name}}
\label{app:responsible_deployment}
{\name} should be deployed with care to avoid false accusations. First, because the MIA score distribution varies slightly across datasets (see Table~\ref{tab:full_mia}), auditors should calibrate the decision threshold on a small set of known non-members from the same domain and select an operating point with a controlled false-positive rate (e.g., FPR$\leq$5\% or even FPR$\leq$1\%). Second, single-query decisions can be noisy; auditors should issue multiple queries and apply the K-means aggregation in Sec.~\ref{sec:pipeline} to retain only high-confidence ones, treating low-agreement cases as inconclusive. Finally, MIA outputs provide statistical evidence rather than definitive proof, and we recommend using {\name} as a screening tool to flag candidates for further investigation rather than as the sole basis for accusation.

\begin{figure*}[t]
    \centering
    \includegraphics[width=0.98\textwidth]{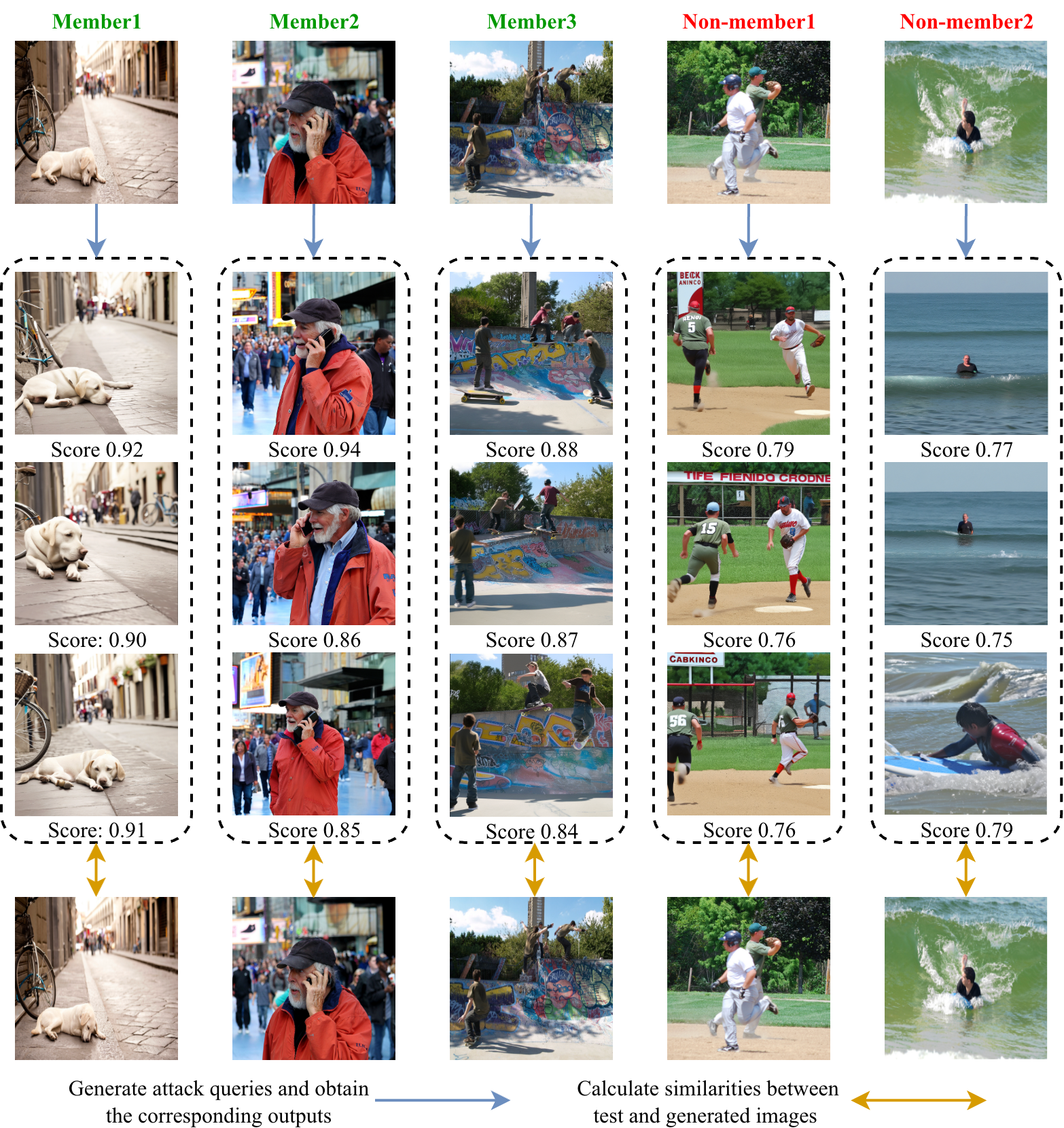}
    \caption{Visual illustration of {\name} on test images from the MSCOCO dataset under {\rag} for the T2I task. Our method performs multiple queries and applies K-means clustering to the resulting MIA scores for selective aggregation.}
    \label{fig:coco_app}
\end{figure*}

\begin{figure*}[t]
    \centering
    \includegraphics[width=0.98\textwidth]{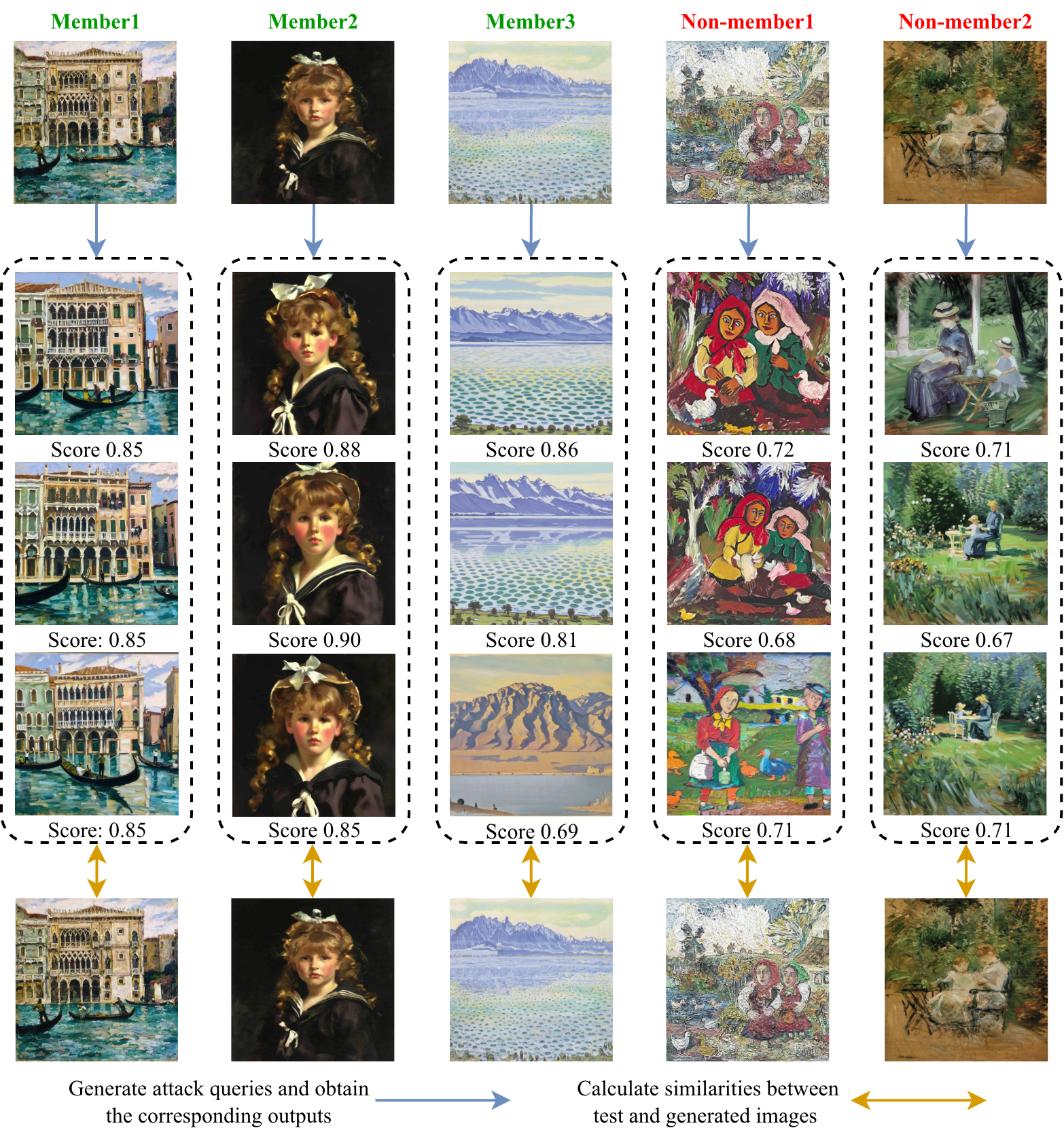}
    \caption{Visual illustration of {\name} on test images from the WikiArt dataset under {\rag} for the T2I task. Our method performs multiple queries and applies K-means clustering to the resulting MIA scores for selective aggregation.}
    \label{fig:wiki_app}
\end{figure*}

\begin{figure*}[t]
    \centering
    \includegraphics[width=0.98\textwidth]{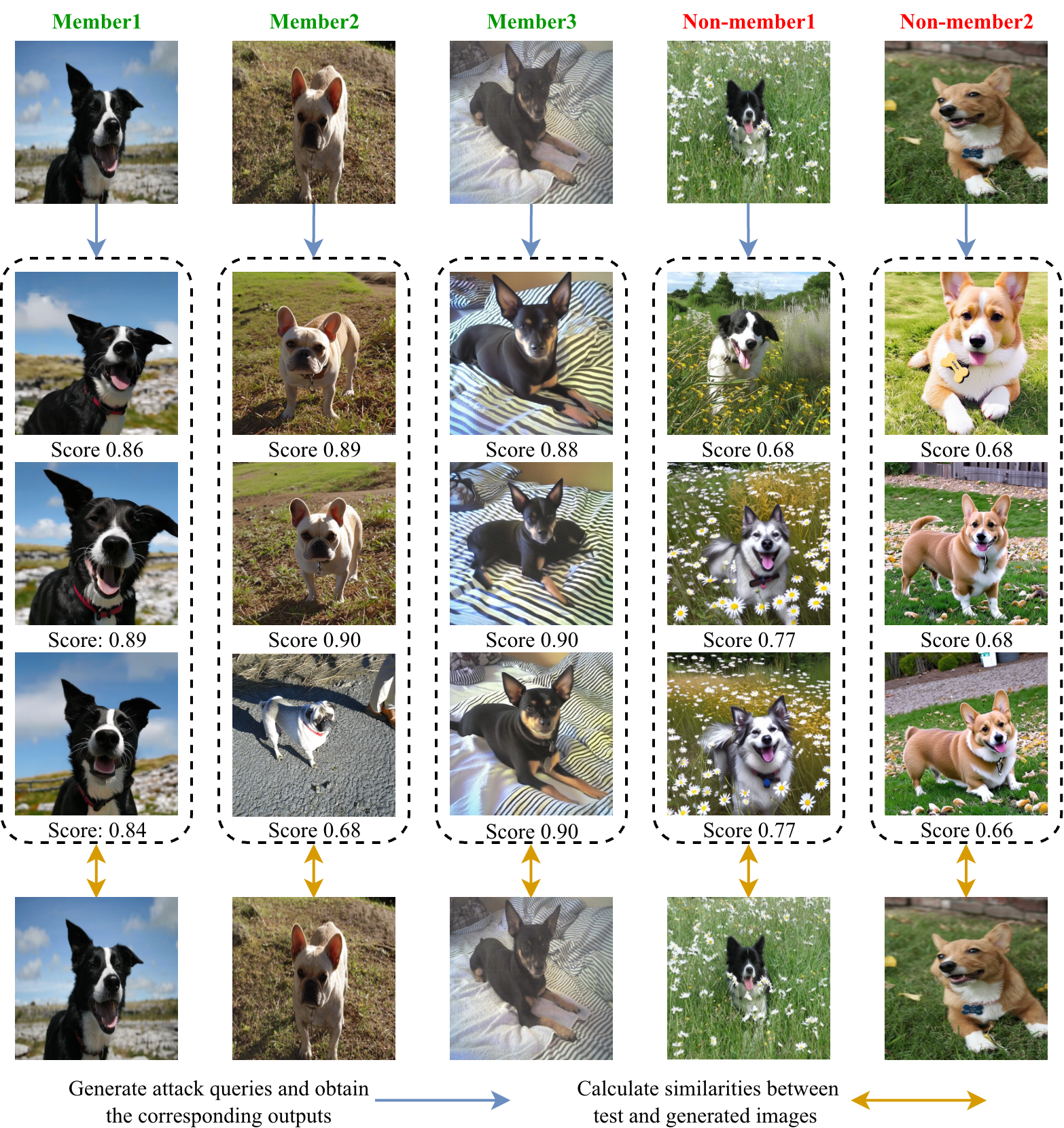}
    \caption{Visual illustration of {\name} on test images from the Stanford Dogs dataset under {\rag} for the T2I task. Our method performs multiple queries and applies K-means clustering to the resulting MIA scores for selective aggregation.}
    \label{fig:dog_app}
\end{figure*}

\section{Qualitative Analysis} 
\label{app:qual}
\begin{figure*}[t]
    \centering
    \includegraphics[width=0.98\textwidth]{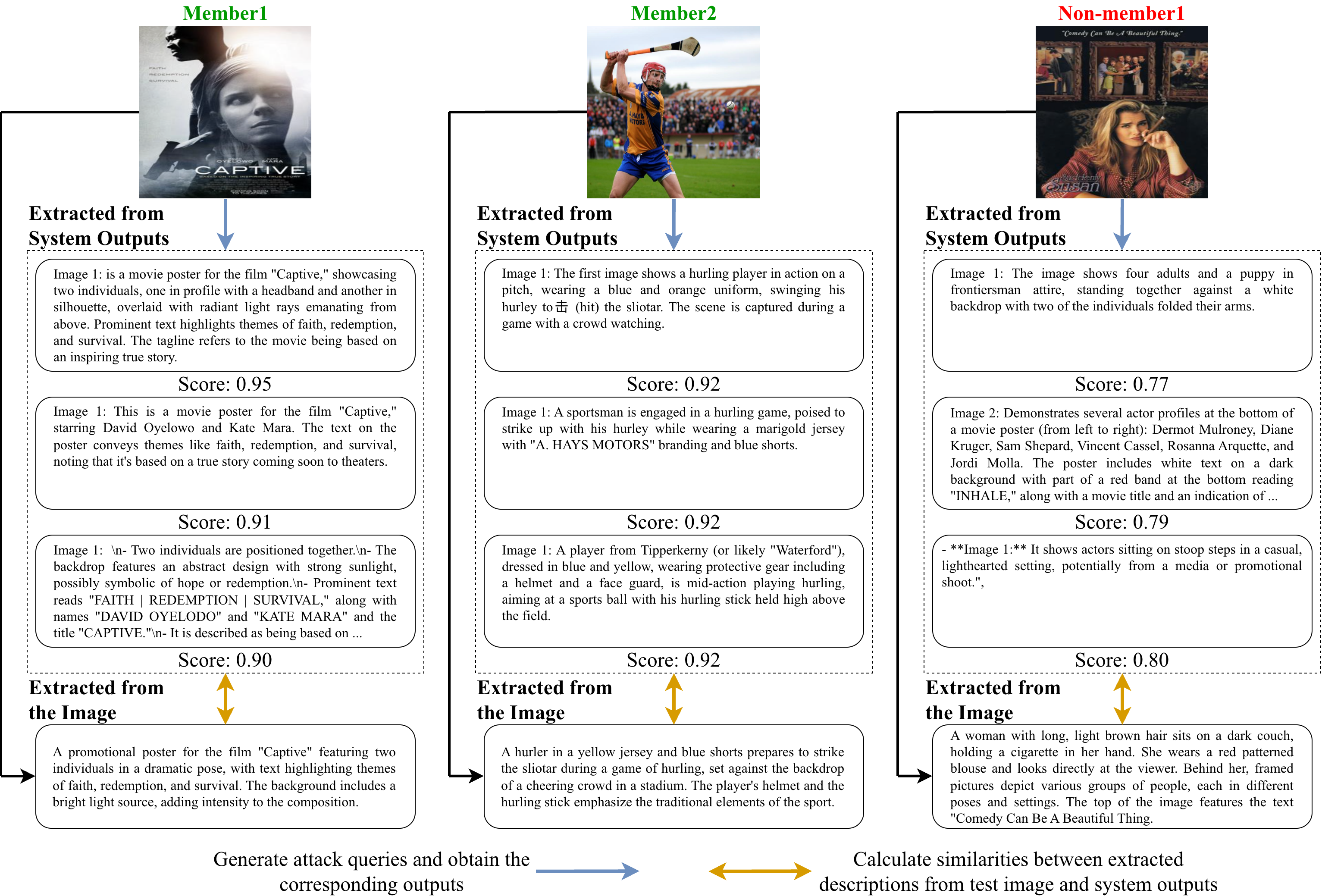}
    \caption{Visual illustration of {\name} on test images from the MMQA dataset under {\rag} for the Q\&A task. Our method performs multiple queries and applies K-means clustering to the resulting MIA scores for selective aggregation.}
    \label{fig:mmqa_app}
\end{figure*}

Figure~\ref{fig:coco_app}--\ref{fig:dog_app} visualize our attack results on the MSCOCO, WikiArt, Dogs datasets under the T2I setting, and Figure~\ref{fig:mmqa_app} visualizes our attack results on the MMQA dataset under the Q\&A setting. Our method effectively distinguishes members from non-members by eliciting distinct generation behaviors. For example, for the member in the first column of Figure~\ref{fig:wiki_app} (a painting of a Venetian palace), \textit{\name} successfully triggers the copy-like behavior of the diffusion model: across all three queries, the generated images closely replicate the reference image's composition, color palette, and architectural details, yielding consistently high similarity scores ($\sim$0.85). In contrast, for the non-member in the last column of Figure~\ref{fig:wiki_app} (a painting of figures in a garden), the generator produces images that share semantic content with the caption (i.e., a mother and daughter sitting in a garden) but differ substantially from the reference in visual composition, lighting, and stylistic details. As a result, the similarity scores remain consistently lower ($\sim$0.67--0.71). This contrast confirms that our caption-based extraction strategy reliably amplifies the membership signal: members trigger near-verbatim reconstruction through the diffusion model's copy-like behavior, while non-members yield only semantically aligned but visually divergent outputs.

\section{Computation Resources}
All experiments were conducted on three single-GPU systems, each equipped with an NVIDIA A6000 GPU with 48GB of VRAM. We did not use multi-GPU training or distributed execution. Both attack optimization and retrieval-augmented generation were performed on a single GPU per run.


\end{document}